\documentclass[12pt]{article}
\usepackage{a4wide}
\usepackage{amssymb}
\usepackage{graphicx}
\usepackage{xcolor}
\begin{document}
{\renewcommand{\thefootnote}{\fnsymbol{footnote}}
\begin{center}
{\LARGE Relational evolution with oscillating clocks}\\
\vspace{1.5em}
Martin Bojowald,\footnote{e-mail address: {\tt bojowald@psu.edu}}
Luis Mart\'{\i}nez\footnote{e-mail address: {\tt lxm471@psu.edu}}
and Garrett Wendel\footnote{e-mail address: {\tt gmw5164@psu.edu}}
\\
\vspace{0.5em}
Institute for Gravitation and the Cosmos,\\
The Pennsylvania State
University,\\
104 Davey Lab, University Park, PA 16802, USA\\
\vspace{1.5em}
\end{center}
}

\setcounter{footnote}{0}

\begin{abstract}
  A fundamental description of time can be consistent not only with the usual
  monotonic behavior but also with a periodic physical clock variable, coupled
  to the degrees of freedom of a system evolving in time. Generically, one
  would in fact expect some kind of oscillating motion of a system that is
  dynamical and interacts with its surroundings, as required for a fundamental
  clock that can be noticed by any other system. Unitary evolution does not
  require a monotonic clock variable and can be achieved more generally by
  formally unwinding the periodic clock movement, keeping track not only of
  the value of the clock variable but also of the number of cycles it has gone
  through at any moment. As a result, the clock is generically in a quantum
  state with a superposition of different clock cycles, a key feature that
  distinguishes oscillating clocks from monotonic time. Because the clock and
  an evolving system have a common conserved energy, the clock is in different
  cycles for different energy eigenstates of the system state.  Coherence
  could therefore be lost faster than observed, for instance if a system that
  would be harmonic in isolation is made anharmonic by interactions with a
  fundamental clock, implying observational bounds on fundamental
  clocks. Numerical computations show that coherence is maintained over long
  time scales provided the clock period is much smaller than the system
  period. A small loss of coherence nevertheless remains and, measured in
  terms of the relative standard deviation of the system period, is
  proportional to the ratio of the system period and the clock period. Since
  the precision of atomic clocks could not be achieved if atomic frequencies
  would be subject to additional variations from coupling to a fundamental
  clock, an upper bound on the clock period can be obtained that turns out to
  be much smaller than currently available direct or indirect measurements of
  time.
\end{abstract}

\newpage

\section{Introduction}

Quantum mechanics does not treat space and time on the same footing, even in
its relativistic version. For instance, a meaningful probabilistic
interpretation of quantum states requires that time evolution is unitary such
that an evolving state is always normalized. Unitarity implies that time will
keep going on forever. Positions in space, by contrast, can easily be limited
to finite regions or periodic boundary conditions, as in the basic examples of
an infinite square well and of a particle required to move on a circle.

Such a distinction between space and time appears to violate relativity, in
particular the general covariance of arbitrary transformations of time and
space coordinates realized in general relativity. In this context, attempts to
combine quantum properties with general relativity in a theory of quantum
gravity have indeed encountered several obstacles collectively referred to as
the problem of time \cite{KucharTime,Isham:Time,AndersonTime}. For instance,
the perpetual nature of time, encoded mathematically in the condition of
unitarity in quantum mechanics, is incompatible with the possibility of the
universe (and therefore time) having a beginning at the big bang, or an end if
the universe happens to collapse in some distant future. Moreover, time in
general relativity is a local coordinate that, in general, need not be defined
in the same way everywhere in space-time even if the universe does not
encounter a physical boundary.

The picture of unitary evolution in which monotonic time seems to be required
to never cease increasing is also add odds with our physical measurements of
time, which are based on periodic phenomena such as planetary orbits, the
moving hands on a clock, or the vibrations of a quartz. (For a detailed
discussion of physical clocks, see for instance \cite{Clocks}.) Time is
measured by periodic processes, but represented mathematically in a linear,
monotonic fashion. There is an interesting dichotomy between periodic and
monotonic phenomena in the context of time, which is hard to resolve because
we have a very intuitive understanding of how we experience time but do not
know well what time is on a fundamental level. We experience time as pointing
from the past to the future without being able to move back, as perhaps
indicated by the second law of thermodynamics. The temporal labels we attach
to events are accordingly based on conventions that imply monotonic behavior:
While the numbers we conventionally assigned to what we call time in a strict
sense (second, minute, hour) reflect the periodic nature of how we measure
brief intervals, the complete date (day, month, year) of an event renders the
assingment monotonic.

The appearance of the traditional periodic processes in measurements of time
is not fundamental but determined by their utility. It is easier to recognize
change by observing a periodic system returning to a fixed state multiple
times, compared with the gradual motion of a monotonic process. In addition,
the compactness of a periodic scale makes it easier to construct portable or
wearable clocks. The prevalence of periodic processes in time measurements
therefore does not imply that time must fundamentally be based on a periodic
process. Then again, the possibility of constructing a monotonic label of time
(and date) by counting the cycles that a hierarchy of periodic processes goes
through, as in our actual time-and-date measurements, shows that the
monotonicity of our experienced time does not imply either that time
fundamentally must be monotonic.

In this situation, any statement about fundamental clocks requires fundamental
physics. Here, we use two main ingredients to argue that time should
fundamentally be based on a periodic process. At the same time, we define what
we mean by a ``periodic process'' in a formulation that does not use a
monotonic background time. First, in order to bring space and time on a more
equal footing in quantum mechanics, a decades-old treatment postulates that
time $t$ should be represented by an operator (or a phase-space coordinate in
a classical theory) just like the spatial position $x$. Such a formulation of
evolution is called relational because it describes how one physical degree of
freedom, $x$, evolves with respect to another physical degree of freedom, $t$,
rather than how a single physical degree of freedom evolves with respect to
some external time parameter. (Of course, in a relational formulation we could
equally well describe how $t$ evolves with respect to $x$, but in keeping with
conventions, we choose to call the reference degree of freedom $t$.) This idea
goes back to Dirac \cite{GenHamDyn1} and Bergmann \cite{BergmannTime} and has
seen much recent interest in the context of combining quantum physics with
relativity and gravity
\cite{GeomObs2,PartialCompleteObs,PartialCompleteObsII,EffTime,EffTimeLong,EffTimeCosmo,MultChoice,ClocksDyn,TwoTimes,BianchiInternal,AlgebraicTime,QuantumRef1,QuantumRef2,QuantumRef3,QuantumRef4,QuantumRef5}.

Secondly, if time as a physical degree of freedom is on an equal footing with
space, or with matter degrees of freedom, it may in general be expected to
have interactions with itself or with other degrees of freedom. In a
fundamental theory, the reference degree of freedom we use to describe time
could be selected from one of the fundamental fields of the standard model of
particle physics, or from a physical degree of freedom that determines the
structure of space-time. If we require that the degree of freedom we call time
is such that it does not have any self-interactions, as in traditional
formulations of relational evolution, this degree of freedom takes on a
special form devised just for the purpose of being able to play the role of
time as we think we know it. Such an assumption would forgo any possibility of
determining fundamental properties of time.  More generally, if time is based
on a fundamental clock, it should generically be expected to have
self-interactions, perhaps described by a potential. The reference clock
degree of freedom could then have ``turning points'' which confine its values
to a certain finite range. At this point we also have to address a common
language problem: Our standard experience of a monotonic background time is so
common that it is built into several physics concepts, such as ``turning
points,'' that we still have to refer to even when we try to formulate an
oscillating fundamental clock.  The conventional term ``turning points''
assumes motion with respect to some background time, but here we are not
interested in this motion. We merely refer to the confinement to a finite
range that may be implied by a potential and, for lack of an alternative term,
use the established dynamical term. Any process with turning points in this
sense will be referred to as ``periodic.''

In a fundamental description, we should therefore be able to make sense of
relational evolution with respect to a confined or periodic reference degree
of freedom. For the sake of clarity, we will reserve the word ``time'' for a
monotonic label in accordance with our common experience of time. A
non-monotonic reference system on which the measure of time is based will be
called a ``clock.'' The distinction between ``time'' and ``clock'' does not
appear in traditional relational evolution, but it is relevant for a
discussion on a fundamental level as presented in this paper.

We will describe and evaluate relational evolution with respect to a periodic
clock degree of freedom in what follows, demonstrating that it is not only
consistent with standard requirements on quantum mechanics but also implies
new and potentially observable effects, as announced in \cite{LocalTime}.

\section{Relational evolution}

Relational evolution as a formal device, used often in quantum gravity and
quantum cosmology, helps to bring time conceptually closer to space by
accompanying the canonical pair of position and momentum, $x$ and $p$, with a
second canonical pair of energy and time, $E$ and $t$. For consistent signs,
not that time $t$ is the momentum of $E$, or $-E$ is the momentum of $t$, as a
consequence of the usual negative sign in the time components of the Minkowski
metric $\eta_{\mu\nu}$. In 4-dimensional notation, one can write these
canonical relationships through a Poisson bracket,
$\{x_{\mu},p_{\nu}\}= \eta_{\mu\nu}$.

The extended description by canonical variables implies that the corresponding
quantum theory should include a time operator $\hat{t}$, in addition to an
energy operator $\hat{E}$, such that $[\hat{t},\hat{E}]=-i\hbar$. The
classical energy equation $E=H(x,p;t)$ with the Hamiltonian $H(x,p;t)$ of the
system, possibly having an explicit time dependence, is then quantized to the
Schr\"odinger equation by representing the energy as a derivative operator
$\hat{E}\psi=i\hbar \partial\psi/\partial t$, acting on wave functions
depending on $x$ and $t$.

In order to make the relationship between time and space more apparent,
we can formulate the energy equation as a constraint,
\begin{equation}
 C_1=-E+H(x,p;t)=0\,,
\end{equation}
moving time and space to the same side of the equation. We initially
introduced a new degree of freedom into the usual formulation, $t$ with
momentum $-E$, and now impose a constraint to make sure that the correct
number of independent parameters is maintained.

In spite of a certain formal semblance between $x$ and $t$ in this
formulation, it does not manage to put space and time on an equal footing.
One remaining difference between these two variables is that the latter's canonical
momentum, $E$, appears linearly in the constraint, while the former's
canonical momentum, $p$, usually appears in a quadratic form. A relativistic
energy equation, such as
\begin{equation} \label{C2free}
 C_2=-E^2+p^2+m^2=0
\end{equation}
 for a free particle with mass $m$,
helps to reduce this difference.

\subsection{Classical formulation}

The reformulation of standard Hamiltonian evolution as a constraint linear in $E$
does not change the assumption that time is monotonic. A constraint generates
equations of motion with respect to an auxiliary parameter $\epsilon$ in the
same form as a Hamilton function generates Hamilton's equations in time. For
instance, the equations generated by $C_1$ are
\begin{eqnarray}
 \frac{{\rm d}x}{{\rm d}\epsilon}&=& \frac{\partial C_1}{\partial
   p}=\frac{\partial H}{\partial p}\,,\\
 \frac{{\rm d}p}{{\rm d}\epsilon}&=& -\frac{\partial C_1}{\partial x}=
 -\frac{\partial H}{\partial x}\,,\\
 \frac{{\rm d}E}{{\rm d}\epsilon}&=& \frac{\partial C_1}{\partial t}=
 \frac{\partial H}{\partial t}\,,\\
 \frac{{\rm d}t}{{\rm d}\epsilon}&=& -\frac{\partial C_1}{\partial
   E}=1\,. \label{dtdl}
\end{eqnarray}
The last equation implies that $t$ is monotonic with respect to $\epsilon$; in
fact, it can be identified with $\epsilon$ up to a constant shift. The
remaining equations then obtain their usual Hamiltonian form, demonstrating
the equivalence of the Hamiltonian and constrained formulations.
Equation~(\ref{dtdl}), derived from the non-relativistic constraint, means
that time in non-relativistic mechanics is, unlike space, inevitably given by
a function that is monotonic in $\epsilon$, and therefore perpetual.

The relativistic constraint introduced so far, $C_2$ in equation
(\ref{C2free}), implies a similar monotonic behavior of time. Initially,
\begin{equation} \label{tepsilon}
 \frac{{\rm  d}t}{{\rm d}\epsilon}= -\frac{\partial C_2}{\partial E}=2E
\end{equation}
is not just a numerical constant. However, the constraint $C_2$ for a free
particle implies, via ${\rm d}E/{\rm d}\epsilon=\partial C_2/\partial t=0$,
that $E$ is constant. The time variable $t$ is therefore still monotonic in
$\epsilon$, although its rate of change, given by $2E$, is no longer universal
but depends on the energy.

This observation shows how we can move closer to a local notion of time based
on a periodic clock: If we find a relativistic model in which there is a
time-dependent potential added to $C_2$, the energy will no longer be
constant. If its value can move through zero and change sign,
${\rm d}t/{\rm d}\epsilon=2E$ would change sign, and $t$ might oscillate for a
suitable potential. These oscillations would be with respect to an external
parameter $\epsilon$, but this parameter is only auxiliary because it can
locally be eliminated from solutions. Yet, in spite of this auxiliary nature
of the parameter in which oscillations may unfold, dynamics with respect to an
oscillating clock degree of freedom representing time would be markedly
different from a monotonic time $t$ because of the presence of a potential,
which we take as the defining feature of an oscillating clock. Recall that
this definition only refers to the form of the constraint and does not require
a background time.

Systems with time-dependent potentials in a relational interpretation
indeed exist in fundamental physics. For instance, the cosmological dynamics
of an expanding universe on large scales is determined by the Friedmann
equation
\begin{equation}
 \left(\frac{1}{a}\frac{{\rm d}a}{{\rm d}t}\right)^2= \frac{8\pi G}{3c^2}\rho\,,
\end{equation}
for the scale factor $a$, whose time derivative is related to the energy
density $\rho$ of matter, Newton's constant $G$, and the speed of light
$c$. It can be formulated as a constraint
\begin{equation} \label{FriedmannConstraint}
C_3= - H(\phi,p_{\phi};V)+\frac{6\pi G}{c^2} Vp_V^2=0
\end{equation}
in canonical variables \cite{Foundations} given by the expanding volume, $V$,
and its momentum, $p_V=-c^2{\cal H}/(4\pi G)$ related to the Hubble parameter
${\cal H}=a^{-1}{\rm d}a/{\rm d}t$. The matter variables are often described
by another canonical pair, $\phi$ and $p_{\phi}$, which appear in the matter
Hamiltonian $H=V\rho$. A common example of an isotropic matter degree of
freedom $\phi$ is a scalar field with mass $m$, in which case the energy
density equals
\begin{equation} \label{rho}
 \rho=\frac{c^2}{2} \frac{p_{\phi}^2}{V^2}+ \frac{1}{2}m^2\phi^2\,.
\end{equation}
For $m=0$, equations of motion imply that $p_{\phi}$ is conserved and $\phi$
is monotonic, much like $E$ and $t$ as determined by the previous constraint,
$C_2$. But the more generic case of $m\not=0$ leads to a representation of
time through a clock degree of freedom $\phi$ that evolves in a non-monotonic,
periodic fashion.

Simplifying some coefficients in the cosmological example, we will now work
with a constraint of the form
\begin{equation} \label{C4}
 C_4= -p_{\phi}^2-\lambda^2 \phi^2+ H(x,p)^2
\end{equation}
with some system Hamiltonian $H(x,p)$, keeping the quadratic dependence on
$\phi$ and $p_{\phi}$ but applying it to non-cosmological models. For
$\lambda=0$, in which case $\phi$ is not confined or periodic, we can
factorize the quadratic constraint into two factors linear in $p_{\phi}$ such
that solutions to the constraint equation $C_4=0$ are equivalent to solutions
of the non-relativistic constraint $C_1=0$ with $p_{\phi}=\pm E$. For
$\lambda\not=0$, $\phi$ is confined for a given system energy, allowing us to
generalize monotonic time behavior to an oscillating clock, $\phi$. Imposing
the constraint $C_4=0$ then couples the system degrees of freedom, $x$ and
$p$, to the clock degrees of freedom, $\phi$ and $p_{\phi}$. Even though there
is no force between system and clock for a constraint of the form (\ref{C4}),
their dynamics are related by the energy-balance constraint $C_4=0$. The
absence of a coupling force relieves us from the obligation to justify any
specific form from fundamental physics. If there were such a coupling force in
addition to the energy-balance constraint, it would only strengthen the
effects of clock-system interactions that we will observe in what follows.

\subsection{Quantum formulation}

Quantum cosmology aims to quantize constraints such as
(\ref{FriedmannConstraint}) or the simpler (\ref{C4}) by solving and
interpreting the quantum constraint equation
$\hat{C}_4\psi(x,\phi)=0$. However, solutions of this equation for the wave
function $\psi$ do not evolve in an obvious way because the imposed quantum
constraint implies that the evolution operator associated with $\hat{C}_4$,
given by $\exp(-i\hat{C}_4\epsilon/\hbar)$ where $\epsilon$ is analogous to
the auxiliary parameter of the same letter used in the classical formulation,
acts trivially on solutions $\psi(x,\phi)$ of $\hat{C}_4\psi(x,\phi)=0$.

\subsubsection{Ordering questions}

A common method to address this problem goes back to Dirac \cite{GenHamDyn1},
called deparameterization and implemented in detail in \cite{Blyth} for
quantum cosmology. This method, which assumes $\lambda=0$ in (\ref{C4}) or
$m=0$ in (\ref{rho}), amounts to an inversion of the process that led us from
Hamiltonians to constraints: We factorize the quantum constraint equation as
\begin{equation} \label{Factor}
\hat{C}_4^{(\lambda=0)}\psi=
  (-\hat{p}_{\phi}^2+ H(\hat{x},\hat{p})^2)\psi=
  (-\hat{p}_{\phi}+H(\hat{x},\hat{p}))(\hat{p}_{\phi}+H(\hat{x},\hat{p}))\psi=0\,,
\end{equation}
such that it can be solved by either of the parentheses being zero when acting
on $\psi$: $\hat{p}_{\phi}\psi=\mp H(\hat{x},\hat{p})\psi$, or
\begin{equation}\label{pmSchroed}
  i\hbar\frac{\partial\psi}{\partial\phi}= \pm H(\hat{x},\hat{p})\psi\,.
\end{equation}
We have obtained Schr\"odinger evolution (for both choices of the orientation
of time) from the quantum constraint.

For $\lambda\not=0$, however, the procedure suggested by Dirac does not go
through precisely because time is no longer monotonic.  Deparameterized
evolution with respect to $\phi$ then cannot be unitary because the classical
$\phi$ oscillates.  A further difficulty appears at a formal level, noting
that the factorization (\ref{Factor}) is not correct if a Hamiltonian
$\hat{\bar{H}}$, such as a quantization of $\bar{H}=
\sqrt{H(x,p)^2-\lambda^2\phi^2}$ which classically solves $C_4=0$ for
$p_{\phi}$, depends on $\phi$. Because $[\hat{p},\hat{\bar{H}}]\not=0$,
\begin{equation} \label{Factor2}
 \hat{C}_4':=(-\hat{p}_{\phi}+\hat{\bar{H}})(\hat{p}_{\phi}+\hat{\bar{H}})=
 -\hat{p}_{\phi}^2-[\hat{p}_{\phi},\hat{\bar{H}}]+ \hat{\bar{H}}{}^2=
 -\hat{p}_{\phi}^2+ \hat{\bar{H}}{}^2+ i\hbar
 \widehat{\frac{\partial\bar{H}}{\partial\phi}}
\end{equation}
does not agree with the constraint $-\hat{p}_{\phi}^2+\hat{\bar{H}}{}^2$; see
also \cite{EffTime,EffTimeLong}.

The commutator term, being proportional to $\hbar$, could be interpreted as a
quantum correction (although a complex-valued one), modifying the classical
constraint $C_4$. The modified constraint equation $\hat{C}_4'\psi=0$ can then
be solved by $-\hat{p}_{\phi}\psi=\hat{\bar{H}}\psi$, using the rightmost
factor in (\ref{Factor2}) next to the wave function $\psi$ in
$\hat{C}_4'\psi=0$.  However, if $\phi$ is a local oscillating clock, both
factors in a version of (\ref{Factor}) are required for forward and backward
evolution with respect to $\phi$. But exchanging the factors in
(\ref{Factor2}), such that $-\hat{p}_{\phi}+\hat{\bar{H}}$ now acts directly
on a wave function $\psi$, modifies the constraint:
\begin{equation} \label{Factor3}
 \hat{C}_4'':=(\hat{p}_{\phi}+\hat{\bar{H}})(-\hat{p}_{\phi}+\hat{\bar{H}})=
 -\hat{p}_{\phi}^2+[\hat{p}_{\phi},\hat{\bar{H}}]+ \hat{\bar{H}}{}^2=
 -\hat{p}_{\phi}^2+ \hat{\bar{H}}{}^2- i\hbar
 \widehat{\frac{\partial\bar{H}}{\partial\phi}} \not=\hat{C}_4'\,.
\end{equation}
It therefore seems impossible to include both signs in a quantized
$p_{\phi}=\pm \bar{H}$ for a unique quantum model, based on a single quantum
constraint.

\subsubsection{Gribov horizons}

In a classical treatment, as $\phi$ evolves through its turning points in the
quadratic potential $\lambda^2\phi^2$, the sign of $p_{\phi}$
alternates. Different half-cycles of this periodic evolution are therefore
governed by not just one but both factors in the classical version of
(\ref{Factor}). The factorization of quadratic quantum constraints appears to
be in conflict with this elementary behavior. However, quantum mechanics is
more subtle. As shown in \cite{Gribov}, the problem of time is a special case
of the Gribov problem of gauge theories \cite{GribovQuant,GribovRev}, where
the $\epsilon$-flow generated by the constraint plays the role of the gauge
flow, and selecting a variable such as $\phi$ as time fixes the gauge as long
as $p_{\phi}\not=0$: Setting $\phi=\tau$ to a constant value $\tau$ of a
global time parameter then gives a cross-section of the flow. When
$p_{\phi}=0$ at a turning point of $\phi$, $\phi$ is at an extremum and the
condition $\phi=\tau$ is not transversal to the flow. Moreover, because $\phi$
is not monotonic, the condition $\phi=\tau$ evaluated on the full evolution
does not have a unique solution.

These issues are common to all gauge theories with Gribov problems, in which
transition surfaces in phase space, such as $p_{\phi}=0$, are called Gribov
horizons. The usual solution to this problem in quantized gauge theories is to
ensure that Gribov horizons are never crossed on a single gauge orbit in order
to avoid double-counting gauge-fixed solutions in a path integral. In the
present case, this means that quantum mechanics cannot allow $p_{\phi}$ to
change sign on a single gauge orbit. However, since our gauge fixing,
$\phi=\tau$, is time-dependent, we may choose a different Gribov region at
different times, such that we choose the region with $p_{\phi}<0$ when $\phi$
moves forward and the region with $p_{\phi}>0$ for backward motion. (According
to (\ref{tepsilon}), forward motion with respect to $\epsilon$ implies $E>0$,
which corresponds to $p_{\phi}<0$.)

In a canonical treatment and with positive $H$, these two regions correspond
to solutions of the constraint annihilated by the left and right factors,
respectively, in (\ref{Factor2}). Acting on wave functions, we therefore seem
subject to the ordering problem, $\hat{C}_4'\not=\hat{C}_4''$, when $\hat{H}$
is $\phi$-dependent. However, when $\phi$ runs backwards ($p_{\phi}>0$),
evolution with respect to $\phi$ is reversed compared with forward motion
($p_{\phi}<0$). Since time reversal in quantum mechanics is associated with
complex or Hermitian conjugation, we may impose the constraint on wave
functions by acting to the left in this case, $\psi\hat{C}_4'=0$, while using
the standard action to the right for forward motion. A single constraint in a
fixed ordering, $\hat{C}_4'$, can then be used to describe both forward and
backward motion of $\phi$. Implementing this concept formally on a Hilbert
space is subtle because, as we will discuss in more detail, turning points
where $p_{\phi}=0$ are energy dependent according to the constraint
equation. (Algebraic formulations of quantum mechanics \cite{LocalQuant}
that generalize Hilbert-space treatments may be useful in this context, as
they turned out to be in other questions about time as well \cite{AlgebraicTime}.)
Time reversals therefore happen at different times for different energy
eigenstates that are superimposed in an evolving wave function. Our specific
constructions will demonstrate that such a formulation is meaningful and
feasible, but we postpone a detailed general discussion of such time-reversal
states to later work.

\subsubsection{Clock and time}

The specific implementation of cycles in which a local clock such as $\phi$
may move forward or backward, solving the problem of oscillating clocks, is
perhaps obvious, with hindsight, but it has been noticed only recently
\cite{Gribov}: We should distinguish carefully not only between background
time $\epsilon$ and a clock variable $\phi$, as formalized by
deparameterization which would also identify $\phi$ with time, but rather
between {\em three} conceptually different notions: background time
$\epsilon$, a clock variable $\phi$, and (as a new ingredient) global
monotonic time $\tau$. The roles of $\phi$ and $\tau$ are indistinguishable in
the usual treatment of deparameterization in which $\phi$ is monotonic and
can be assumed to be identical with $\tau$. If $\phi$ has turning points,
however, it can be identified with a linear function of $\tau$ only locally,
for periods of evolution that do not contain a turning point of $\phi$. After
a turning point, we should realign the relationship between $\phi$ and $\tau$
such that $\tau$ keeps on going forward while $\phi$ moves back. For instance,
if $\phi(\tau)=\tau+A$ before a turning point at $\tau_{\rm t}$ (with some
constant $A$), $\phi(\tau)=-\tau+A+2\tau_{\rm t}$ rewinds $\phi$ in a way that
is connected continuously to $\phi(\tau)$ before the turning point.

The new distinction between three types of variables related to time has
conceptual implications that will not be the focus of this paper, and which
we mention only briefly in this paragraph. For instance, there may be no time
operator because time $\tau$, as outlined below, is a constructed, effective
parameter and not fundamental. There would only be a clock operator
$\hat{\phi}$ which can measure the direction in which $\phi$ points in its
cycle, but not which cycle it is in. The number of cycles, and therefore time,
would have to be determined by keeping track of a suitable succession of clock
measurements. It would not be possible to determine time by a single
measurement because, unlike the clock, it is not represented by a fundamental
degree of freedom. As we will also see, a quantum clock is generically in a
superposition of different cycles, such that the cycle is not a sharply
defined observable.

Since classical physics, not restricted by the condition of unitarity, can
easily be formulated with local times, we usually do not have to introduce a
time parameter such as $\tau$. In this context, the transition from a
background parameter $\epsilon$ to a clock variable $\phi$ is often motivated
as moving a step closer toward a fundamental description of time, no longer
given by a mathematical coordinate but rather by a physical measurement by
means of a clock. The value of $\phi$ is then the position of a periodic
phenomenon used as a clock, modeled in our example with regular periods by the
harmonic-oscillator Hamiltonian $p_{\phi}^2+\lambda^2\phi^2$ added to our
constraints. However, observing $\phi$ no longer corresponds to a physical
measurement of time. It is merely the reading of a clock instant without putting
it into the context of a constructed time-and-date label of events. The
variable $\phi$, like a Cartesian coordinate used to determine the position of
the hands on a clock, oscillates back and forth, but the time we infer from
this motion always increases. This perpetually increasing time is the global
time $\tau$ introduced here.

Because ${\rm d}\phi/{\rm d}\tau$ changes sign at the turning points of
$\phi$, a parameterization with respect to global time is consistent with
having alternating signs of $p_{\phi}$ (related to our specification of Gribov
regions) even while the energy (of a stable system) should always be positive:
Building on (\ref{pmSchroed}), the Schr\"odinger equation
\begin{equation} \label{Schroedingertau}
  i\hbar\frac{\partial\psi}{\partial\tau}= i\hbar\frac{{\rm d}\phi}{{\rm
      d}\tau} \frac{\partial\psi}{\partial\phi}= -\frac{{\rm
      d}\phi}{{\rm d}\tau} {\rm sgn}(\hat{p}_{\phi})
  H(\hat{x},\hat{p},\phi(\tau))\psi  
=H(\hat{x},\hat{p},\phi(\tau))\psi
\end{equation}
with respect to global time $\tau$ contains both branches of wave-function
evolution with $\hat{p}_{\phi}=\mp H(\hat{x},\hat{p})$, even if the
Hamiltonian $\hat{H}$ is always positive. Since we have forward motion (now
with respect to our new parameter $\tau$) for $p_{\phi}<0$ and backward motion
for $p_{\phi}>0$, we always obey the condition
$-({\rm d}\phi/{\rm d}\tau){\rm
  sgn}(p_{\phi})=1$. Equation~(\ref{Schroedingertau}) is therefore the
standard Schr\"odinger equation with a positive Hamiltonian.  Global, 
unitary time evolution with respect to $\tau$ is then defined by taking
into account the sign changes of $p_{\phi}$ in strict correspondence with the
sign changes of ${\rm d}\phi/{\rm d}\tau$ for forward and backward motion of
$\phi(\tau)$.

Still, even though (\ref{Schroedingertau}) looks like a standard Schr\"odinger
equation, it depends on a piecewise linear but not strictly linear
parameterization $\phi(\tau)$ and therefore implies new features. At turning
points of $\phi$, the dependence of $\phi$ on $\tau$ changes abruptly, which
can be implemented in solutions by concatenating evolution operators
$\exp(-i\int \hat{H}{\rm d}\tau/\hbar)$ derived for strictly linear branches
of $\phi(\tau)$, or their transition amplitudes. The sudden changes in a wave
function constructed from such concatenated evolution operators imply that the
time-dependent phase lacks smoothness, but continuity and unitarity are never
compromised.

While this procedure introduced in \cite{Gribov} presents a solution to some
aspects of the problem of time, the first to make sense of a local notion of
an oscillating clock variable $\phi$, the question of its physical viability
remained open. For instance, one might worry that the rather sudden changes of
$\tau$-evolution operators at turning points, and correspondingly of the phase
of the wave function, could destroy coherence faster than in standard quantum
mechanics with a background time. They might then be in conflict with
sensitive experimental observations, for instance in atomic clocks.  In
particular, the turning-point condition $p_{\phi}=0$ is met at different clock
values $\phi$ for different system energies because of the constraint, and
therefore at different global times $\tau$. The independent energy
contributions in a coherent state would therefore be affected differently by
turning points, endangering their delicate balance required for long-term
coherence.

In the next section we will show that this concern is unwarranted: Coherence
remains intact over long time scales, provided the fundamental clock is
sufficiently fast, with durations of cycles much shorter than the typical rate
of change of the non-time observable, $x$. The procedure of oscillating clocks
is therefore physically viable, and it is testable by measurements of quantum
coherence. Detailed calculations presented in what follows impose a tight
upper bound on the possible fundamental period of time.

\section{Global evolution}

Our formal results are valid for a constraint of the form (\ref{C4}) with an
arbitrary Hamiltonian $H(x,p)$ of a bound-state system. Since any initial
state can be written as a superposition of eigenstates of
$\hat{H}=H(\hat{x},\hat{p})$, it is sufficient to compute evolving wave
functions or transition amplitudes by solving the ordinary differential
equations
\begin{equation} \label{SchroedingerE}
i\hbar \frac{{\rm d}\psi_k(\phi)}{{\rm d}\phi}=\pm
\sqrt{E_k^2 -\lambda^2\phi^2} \ \psi_k(\phi)\,,    
\end{equation}
in the energy representation, where $E_k$ is one of the energy eigenvalues of
$\hat{H}$ and $\psi_k$ the corresponding eigenfunction. This differential
equation is straightforward to solve, giving
\begin{equation} \label{psik}
        \psi_k(\phi) =\psi_k (0)
        \exp \left( \mp\frac{i }{2 \hbar} \left(
            \phi
            \sqrt{E_k^2-\lambda^2\phi^2}+\frac{E_k^2}{\lambda}
            \arcsin\left(\frac{\lambda\phi}{E_k}\right) \right)
        \right)\,.
\end{equation}
(Without loss of generality, we assume that the initial phase of
$\psi_k(\phi)$ with respect to the energy eigenstate $\psi_k$ vanishes.)
As a function of the energy eigenvalues, the phase function
\begin{equation} \label{Theta}
 \Theta_k(\phi)= -\frac{1}{2 \hbar} \left(
            \phi
            \sqrt{E_k^2-\lambda^2\phi^2}+\frac{E_k^2}{\lambda}
            \arcsin\left(\frac{\lambda\phi}{E_k}\right)\right)
\end{equation}
can be used for any bound-state system, provided the clock Hamiltonian is
given by $p_{\phi}^2+\lambda^2\phi^2$.

\begin{figure}
    \centering
    \includegraphics[width=0.6\textwidth]{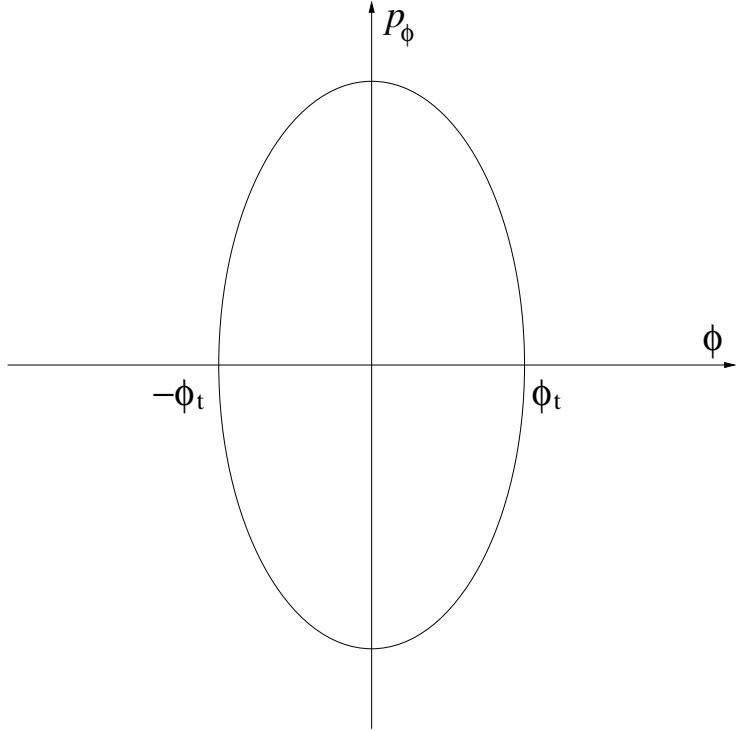}
    \caption{Harmonic clock degree of freedom in phase
      space. \label{Fig:Phipphi}}
\end{figure}

When $E_k^2 < \lambda^2\phi^2$, Eq.~(\ref{psik}), taken at face value,
produces a wave function that is not normalized, highlighting the unitarity
problem of deparameterization with an oscillating clock. Unitarity starts
being violated precisely when $\phi=\pm\phi_{\rm t}$, where
\begin{equation} \label{phit}
    \phi_{\rm t}=\frac{E_k}{\lambda}\,,
\end{equation}
reaches a turning point corresponding to the energy $E_k$. For later
reference, we illustrate the phase-space trajectory of the clock degree of
freedom in phase space in Fig.~\ref{Fig:Phipphi}. For a complete cycle, we
clearly need both positive and negative $p_{\phi}$.

\subsection{Unwinding time}

\begin{figure}
    \centering
    \includegraphics[width=0.8\textwidth]{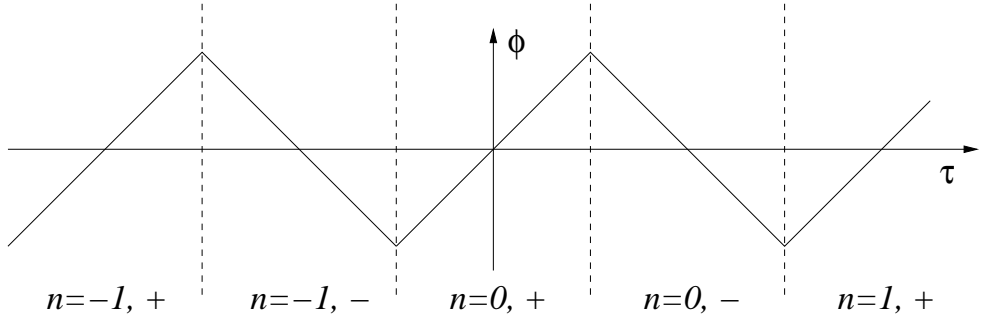}
    \caption{A periodic clock degree of freedom $\phi$ as a function of global
      monotonic time, as constructed in Eq.~(\ref{phitau}). The parameter $n$
      counts the clock cycle according to Eq.~(\ref{n}), starting at $\tau=0$,
      while $\pm$ indicate the sign of ${\rm d}\phi/{\rm d}\tau$, determining
      which one of the two options (\ref{Theta1}) and (\ref{Theta2}) should be
      used at a given $\tau$. \label{Fig:phitau}}
\end{figure}

In order to solve the unitarity and sign problems, following \cite{Gribov}, we
introduce a monotonic global time $\tau$ related to the clock variable $\phi$
in a continuous and piecewise linear fashion:
\begin{equation} \label{phitau}
\phi(\tau)=\left\{\begin{array}{cl}
      \tau -4n\phi_{\rm t} & \mbox{if }\quad 4n-1 \leq \tau/\phi_{\rm t} \leq
      4n+1\\
(4n+2)\phi_{\rm t}-\tau & \mbox{if }\quad 4n+1 \leq \tau/\phi_{\rm t} \leq
      4n+3
       \end{array}\right.\,.
\end{equation}
Here, the integer
\begin{equation} \label{n}
    n =\left \lfloor \frac{1+\tau/\phi_{\rm t}}{4} \right \rfloor
\end{equation}
equals the number of clock cycles, starting with $n=0$ at $\tau=0$. This
parameterization, illustrated in Fig.~\ref{Fig:phitau}, is constructed such
that $\phi$ (i) is related to $\tau$ in a piecewise linear fashion with equal
rates for $\phi$ and $\tau$ (${\rm d}\phi/{\rm d}\tau=\pm 1$), and (ii) never
takes values outside of the range delimited by its turning points,
$\pm\phi_{\rm t}$. Time $\tau$ therefore progresses at the same rate as the
clock, and it keeps track of the number of clock cycles that have passed while
it unwinds the periodic behavior of the clock.

Implicitly, each energy eigenstate contained in a system state dictates its
own clock period $4\phi_{\rm t}$ through the $\phi_{\rm t}$-dependence in
$\phi(\tau)$, where $\phi_{\rm t}$ depends on $E_k$ according to
(\ref{phit}). For simplicity, we dropped the subscript $k$ in the more
complete notation $\phi_{k}(\tau)$ because we will for some time be working
with individual energy eigenstates. However, when we bring different energy
eigenstates back in superposition, the $E_k$-dependence of $\phi_{\rm t}$
implies that a unique global time $\tau$ for the entire state requires
different energy eigenstates to be at different clock values $\phi$ and in
different cycles. The combined clock-system state therefore evolves into a
superposition of different clock cycles whenever the system is in a
superposition of different energy eigenstates.

Inserting $\phi(\tau)$ in
$\psi_k(\phi)=\psi_k(\phi_0)\exp(\pm i(\Theta_k(\phi)-\Theta_k(\phi_0)))$,
with some initial $\phi_0$ in a given half-cycle of the clock, results in the
local solutions
\begin{equation}\label{Theta1}
 \psi_k(\tau)= \psi_k(\phi_0) \exp(\pm i(\Theta_k(\phi(\tau))-\Theta_k(\phi_0)))
\end{equation}
of (\ref{Schroedingertau}). In any given half-cycle of the clock, the sign in
the exponent has to be chosen such that it cancels the sign
${\rm d}\phi/{\rm d}\tau$ produced by acting on $\Theta_k(\phi(\tau))$ with a
$\tau$-derivative in the Schr\"odinger equation, using the chain rule. This
condition, as introduced before, ensures that $\tau$-evolution is generated by
a positive Hamiltonian for a stable system.

Therefore,
\begin{equation}
 \psi_k(\tau)= \psi_k(\tau_0) \exp( i{\rm sgn}({\rm d}\phi/{\rm d}\tau)
 (\Theta_k(\phi(\tau))-\Theta_k(\phi(\tau_0))))
\end{equation}
or, in a piecewise description,
\begin{equation}\label{Theta1}
 \psi_k(\tau)= \psi_k(\tau_1) \exp\left( i( \Theta_k(\phi(\tau))-
 \Theta_k(\phi(\tau_1)))\right)  
\end{equation}
during any half-cycle with ${\rm d}\phi/{\rm d}\tau>0$ (starting at some
$\tau_1$), while
\begin{equation}\label{Theta2}
 \psi_k(\tau)=   \psi_k(\tau_2) \exp\left( i( \Theta_k(\phi(\tau))-
 \Theta_k(\phi(\tau_2)))\right)
\end{equation}
when ${\rm d}\phi/{\rm d}\tau<0$ ({\em ending} at some $\tau_2$). The latter
equation implies that
\begin{equation} \label{Theta2p}
 \psi_k(\tau_2)=   \psi_k(\tau) \exp\left(- i( \Theta_k(\phi(\tau))-
 \Theta_k(\phi(\tau_2)))\right)
\end{equation}
with the opposite sign in the phase for forward evolution in a backward
half-cycle, as required. This equation may be used whenever $\tau_2>\tau$ in
the same backward half-cycle.

The concatenated solution then has a continuous phase because the phases of
the two wave functions (\ref{Theta1}) and (\ref{Theta2}) indeed meet in the
middle: At a turning point, if $\phi_k(\tau_1)$ marks the beginning of a
forward half-cycle and $\phi_k(\tau_2)$ marks the end of the next backward
half-cycle, we may choose $\tau$ to be in both half-cycles, interpreted either
as the end of the first one or the beginning of the second
one. Correspondingly, the phase $\Theta_k(\phi(\tau))- \Theta_k(\phi(\tau_1))$
added to $\psi_k(\tau_1)$ and the phase
$\Theta_k(\phi(\tau))- \Theta_k(\phi(\tau_2))$ added to $\psi_k(\tau_2)$
produce the same state. This solution is globally valid because $|\phi(\tau)|$
never surpasses $\phi_{\rm t}$, demonstrating unitarity. While phase
continuity is guaranteed by construction, smoothness or even differentiability
is not. The system maintains unitary evolution as the wavefunction itself is
smooth during any half-cycle, whereas different evolution operators are
concatenated (rather than extended by solving a single differential equation)
precisely where the phase is not differentiable.

In our specific example, according to the phase $\Theta_k(\phi(\tau))$ in
(\ref{psik}), each half-clock cycle of $\phi$, changing monotonically from
$-\phi_{\rm t}$ to $\phi_{\rm t}$ or back, adds an amount of
\begin{equation} \label{DeltaTheta}
 \frac{1}{2}\Delta\Theta_k=\Theta_k(\phi_{\rm t})-\Theta_k(-\phi_{\rm t})=
 -\frac{E_k^2}{2\lambda\hbar}
 \left(\arcsin(1)-\arcsin(-1)\right)= -\frac{\pi E_k^2}{2\lambda\hbar}
\end{equation}
to the phase. A forward half-cycle starts at $-\phi_{\rm t}$ and ends at
$\phi_t$ and has a phase changing according to $\Theta_k(\phi)$, while a
backward half-cycle starts at $\phi_{\rm t}$ and ends at $-\phi_{\rm t}$ but
has a phase changing according to $-\Theta_k(\phi)$. Therefore, in both cases
the phase added per half-cycle equals
$\Theta_k(\phi_{\rm t})-\Theta_k(-\phi_{\rm t})$.  A full clock cycle, going
from $-\phi_{\rm t}$ to $\phi_{\rm t}$ and back, adds twice this phase. Going
back in $\phi$ does not cancel out the phase of the previous half cycle
because of our specific construction in which we flip the sign of $\Theta_k$
according to the sign of ${\rm d}\phi/{\rm d}\tau$, dictated by positivity of
the Hamiltonian for a stable system.

As an example, starting at the beginning of the zeroth cycle according to
Fig.~\ref{Fig:phitau}, such that $\tau_0=-\phi_{\rm t}$, $\tau$ increases from
$-\phi_{\rm t}$ to $\phi_{\rm t}$ during the first monotonic phase of $\phi$,
where the latter also increases from $-\phi_{\rm t}$ to $\phi_{\rm t}$. The
next monotonic phase then completes the zeroth cycle and has $\tau$ increasing
from $\phi_{\rm t}$ to $3\phi_{\rm t}$ while $\phi$ decreases back to its
initial value, $-\phi_{\rm t}$. The next clock cycles ($n=1,2,\ldots$) repeat
this process.  Since a full clock cycle adds $-\pi E_k^2/(\lambda\hbar)$ to
the phase, according to (\ref{DeltaTheta}), at the end of the $n$th cycle the
wave function equals
\begin{equation}
 \psi_k(\tau_1)=\psi_k(\tau_0)
        \exp \left( \frac{-i n\pi   E_k^2}{\lambda \hbar} \right)
\end{equation}
where $\phi(\tau_1)=-\phi_{\rm t}=\tau_0$.
It then proceeds by
\begin{eqnarray} \label{psik1}
        \psi_k(\tau) &=& \psi_k(\tau_0)
        \exp \left( \frac{-i n\pi   E_k^2}{\lambda \hbar} \right)
        \exp\left(i(\Theta_k(\phi(\tau))-\Theta_k(-\phi_{\rm
            t}))\right)\nonumber\\
&=& \psi_k(\tau_0)
        \exp \left( \frac{-i (n+1/4)\pi E_k^2}{\lambda \hbar} \right)
        \exp(i\Theta_k(\phi(\tau)))
\end{eqnarray}
as long as $4n-1 \leq \tau/\phi_{\rm t} \leq 4n+1$ ($\phi(\tau)$
increasing). This half-cycle, ending at a time $\tau_2'$ when
\begin{equation}
 \psi_k(\tau_2')= \psi_k(\tau_0) \exp \left( \frac{-i  (n+1/4)\pi E_k^2}{\lambda
     \hbar} \right)
        \exp(i\Theta_k(\phi_{\rm t})) = \psi_k(\tau_0) \exp \left( \frac{-i 
            (n+1/2) \pi E_k^2}{\lambda
     \hbar} \right)
\end{equation}
adds an additional $-\pi E_k^2/(2\lambda\hbar)$ to the phase.  According to
(\ref{Theta2p}), the wave function in the next half-cycle is then given by
\begin{eqnarray}\label{psik2}
        \psi_k(\tau) &=& \psi_k(\tau_0)
        \exp \left( \frac{-i (n+1/2)\pi E_k^2}{\lambda \hbar} \right)
        \exp\left(-i(\Theta_k(\phi(\tau))-\Theta_k(\phi_{\rm
            t}))\right)\nonumber\\
 &=&  \psi_k(\tau_0)
        \exp \left( \frac{-i (n+3/4)\pi E_k^2}{\lambda \hbar} \right)
        \exp(-i\Theta_k(\phi(\tau)))
\end{eqnarray}
($4n+1 \leq \tau/\phi_{\rm t} \leq 4n+3$, $\phi(\tau)$ decreasing). Because
this half-cycle ends when $\phi(\tau)=-\phi_{\rm t}$, the final phase increase
during the combination of two half-cycles equals $-\pi E_k^2/(\lambda\hbar)$,
in agreement with (\ref{DeltaTheta}).

\subsection{Harmonic oscillators}

\begin{figure}
    \centering
    \includegraphics[width=0.7\textwidth]{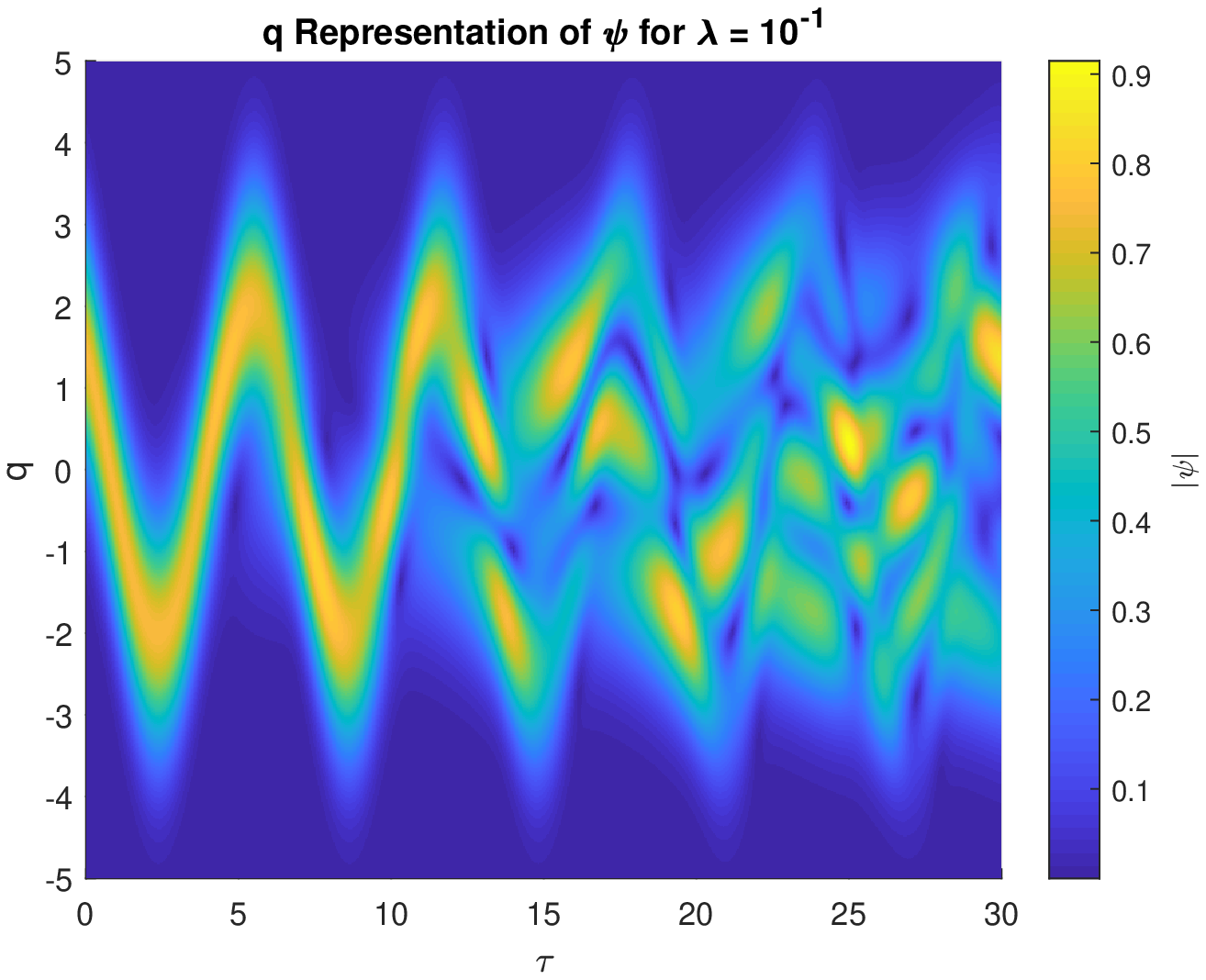}\\
\includegraphics[width=0.7\textwidth]{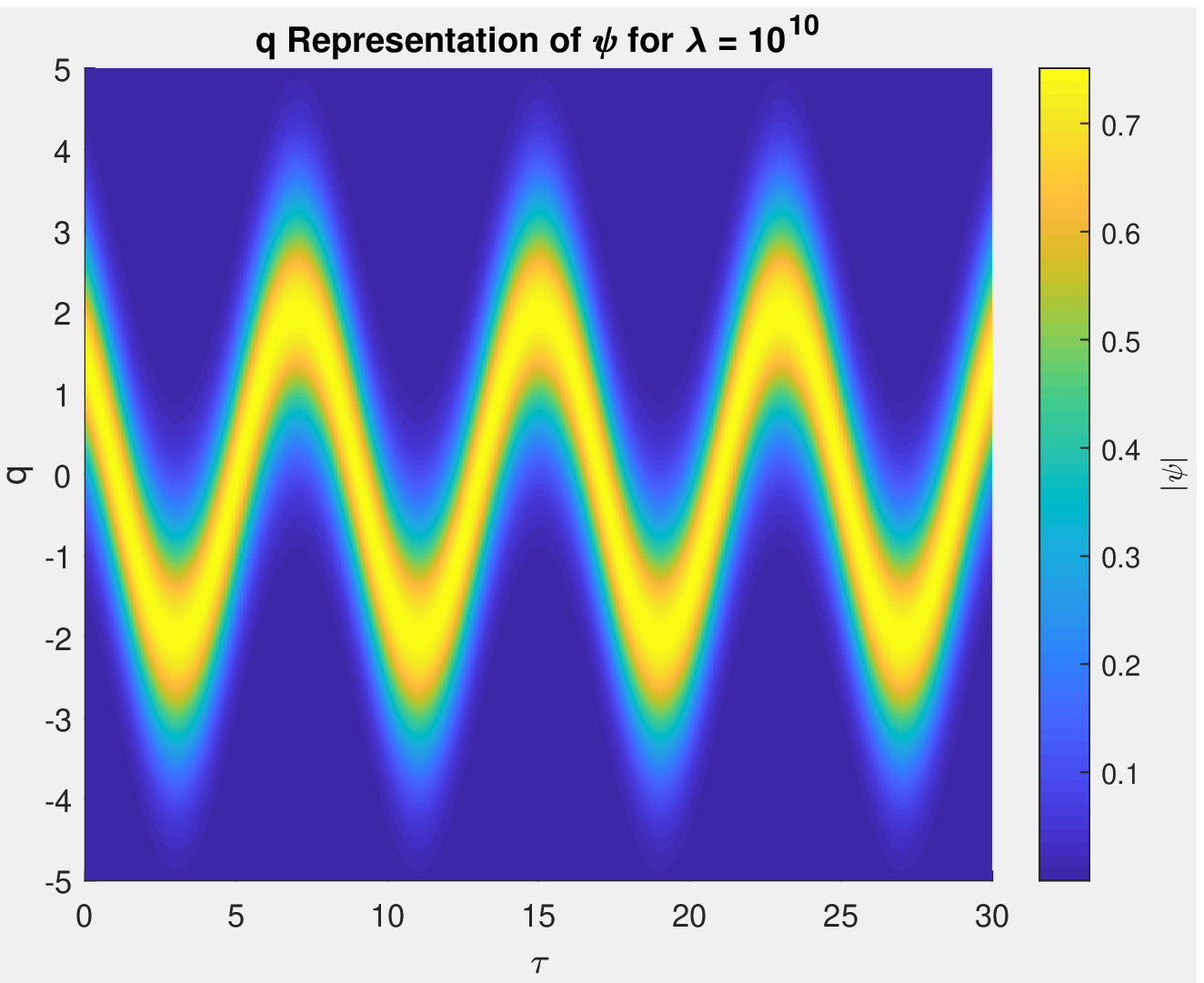}
\caption{Density plot of the wave function for intermediate (top) and large
  $\lambda$ (bottom), using a harmonic-oscillator Hamiltonian
  $\hat{H}=\frac{1}{2}(\hat{p}^2+\hat{x}^2)$ and a standard coherent initial
  state. Coherence is quickly lost in $\tau$-evolution for intermediate
  $\lambda$ as soon as the first turning points are reached (around
  $\tau\sim 10$ for $\lambda=10^{-1}$, top). By contrast, coherence is
  maintained for long times for large $\lambda$ even though billions of
  turning points are crossed during a single system period for
  $\lambda=10^{10}$ (bottom). Strong coherence is maintained in this case even
  though the actual $\phi$-Hamiltonian $(\hat{H}^2-\lambda^2\phi^2)^{1/2}$ is
  not harmonic if $\lambda\not=0$. \label{Fig:psilam}}
\end{figure}

In order to highlight features of coherence, we now consider the example of a
harmonic system Hamiltonian. Since there is perfect coherence in suitable
states of the standard harmonic oscillator, any loss of coherence implied by
our treatment of local clocks can easily be discerned.
Section~\ref{s:NonHarm} will show analogous properties in non-harmonic
examples in order to confirm the general qualitative features.

The parameter $\lambda$ in the harmonic clock Hamiltonian $p_{\phi}^2+\lambda^2\phi^2$
acts as a scaling factor for the frequency of the clock variable. It does not
have the units of frequency but rather of frequency times energy. In fact, the
number of clock cycles in any range of global time depends, through the value
$\phi_{\rm t}=E_k/\lambda$, on the energy of the system, or rather on the
contributions of energy eigenstates to an evolving system state. Since
$\lambda$ determines how often the clock turns around, it must therefore refer
to the energy.

\begin{figure}
    \centering
    \includegraphics[width=0.7\textwidth]{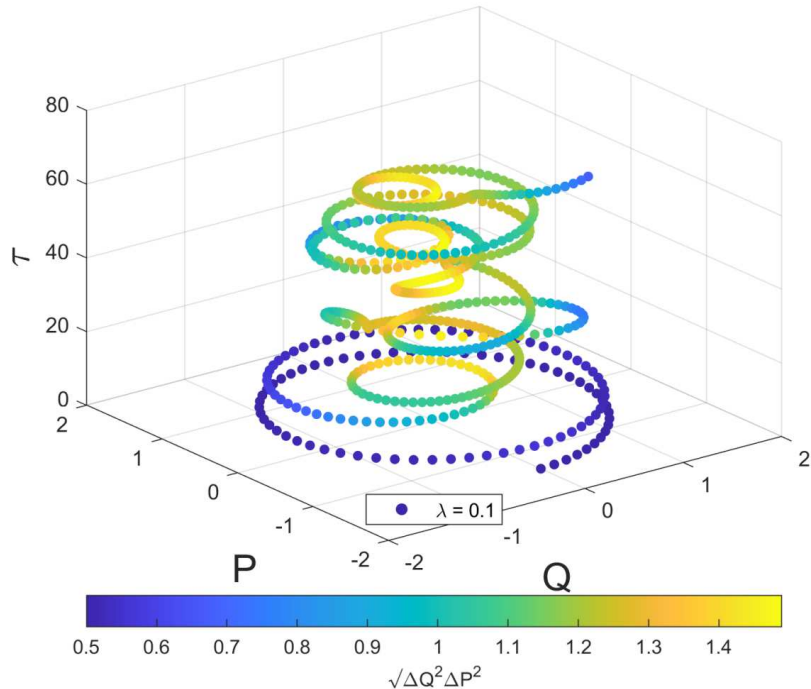}
    \caption{Phase-space trajectory and quantum uncertainty (bottom bar) for a
      harmonic system Hamiltonian with intermediate
      $\lambda=0.1$. \label{Fig:PSLambda01}}
\end{figure}

\begin{figure}
    \centering
    \includegraphics[width=0.7\textwidth]{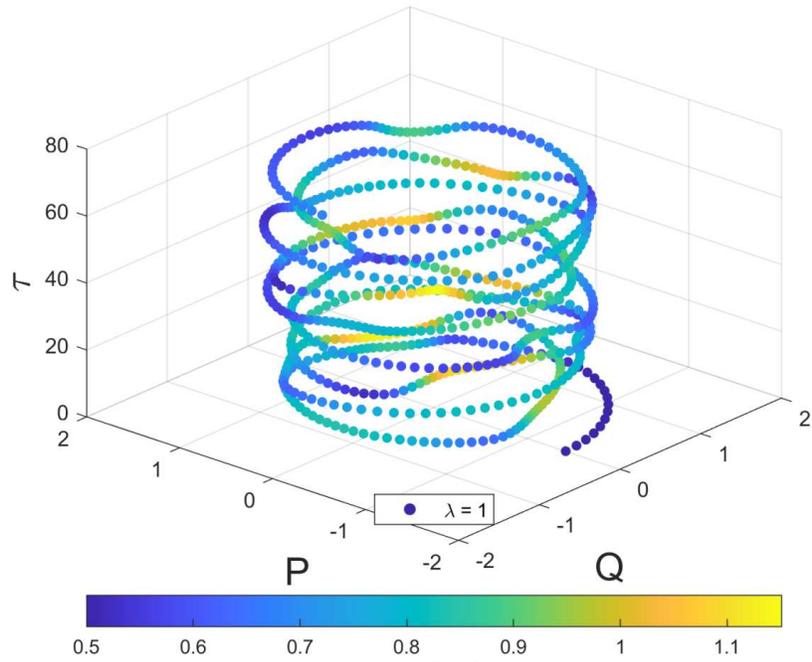}\\[-10mm]
    \includegraphics[width=0.7\textwidth]{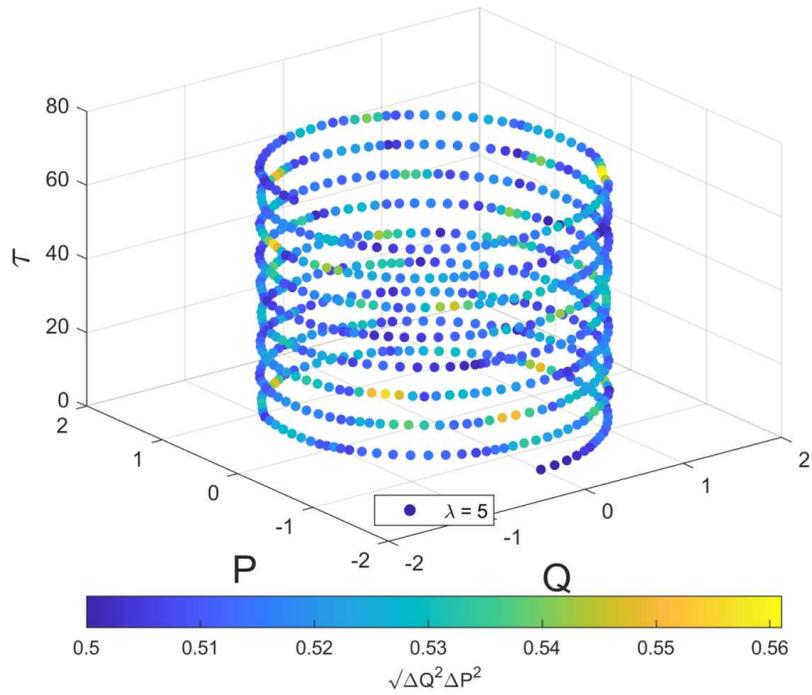}
    \caption{Phase-space trajectory and quantum uncertainty (bottom bars) for a
      harmonic system Hamiltonian. Compared with Fig.~\ref{Fig:PSLambda01},
      coherence and semiclassical behavior are regained as
      $\lambda$ is increased, here showing the examples of $\lambda=1$ and
      $\lambda=5$. \label{Fig:PSLambda1}}
\end{figure}

The parameter $\lambda$ directly affects how often an energy eigenstate (of
the system Hamiltonian) moves through turning points of $\phi$ in a given
$\tau$-interval. As usual, any initial state can be expanded as a
superposition of energy eigenstates, but since different eigenstates imply
different times for the turning points of a clock, an oscillating clock makes
the superposition evolve in a modified way compared with absolute time in
standard quantum mechanics. Figures~\ref{Fig:psilam} and \ref{Fig:PSLambda01}
show how complicated this new dynamics can be, even for a coherent initial
state of a standard harmonic system Hamiltonian.  However, the behavior
simplifies not only when $\lambda \rightarrow 0$, in which case we have
standard quantum mechanics with a monotonic time, but also, surprisingly, when
$\lambda \rightarrow \infty$ as evidenced by Fig.~\ref{Fig:psilam}. The
different behaviors of quantum fluctuations and deviations of the expectation
values from classical sinusoidal behavior are also illustrated in
Figs.~\ref{Fig:PSLambda01} and \ref{Fig:PSLambda1}.

\subsubsection{Small-$\lambda$ approximation}

As $\lambda \rightarrow 0$, the state never approaches a turning point in any
finite range of $\tau$. Far away from these turning points, the phase of the
state resembles that of an unconstrained system, solving
(\ref{SchroedingerE}). Indeed, rewriting the constraint equation (\ref{C4}) by
substituting $\lambda \approx 0$ yields the relation $\hat{p}_\phi \approx
\pm\hat{H}$, belonging to a system with global
time $\phi$.

Utilizing this approximation, the equation of motion is equivalent to the
familiar Schr\"odinger equation
\[
i\hbar \frac{\partial\psi_k(\phi)}{\partial\phi}=
\hat{H} \psi_k(\phi) = E_k\psi_k(\phi)
\]
(choosing a sign suitable for stability from a positive Hamiltonian).  This
equation generates the expected global time-dependent phase for a stationary
state,
\begin{equation}
         \psi_k(\tau) = \psi_k(0)
        \exp \left( \frac{-i \tau E_k }{ \hbar} \right)
\end{equation}
if we identify clock $\phi$ and time $\tau$ in this case.

\subsubsection{Large-$\lambda$ approximation}
\label{s:Largel}

\begin{figure}
    \centering
    \includegraphics[width=0.8\textwidth]{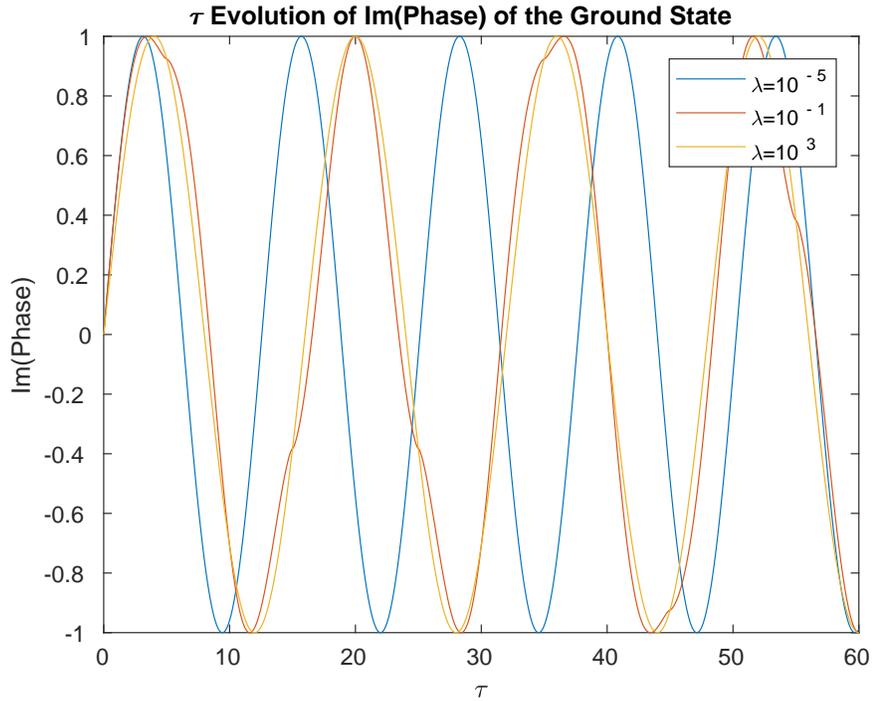}
    \caption{Phases of the ground state of the harmonic potential
      for different values of $\lambda$. Small plateaus caused by individual
      turning points of $\phi$ are clearly visible for intermediate
      $\lambda=10^{-1}$. These plateaus stretch out the curve in the time
      direction. As a consequence, the system period for a large
      $\lambda=10^3$ is greater than the period for a small $\lambda=10^{-5}$.
      Plateaus are no longer visible for large $\lambda$ because the clock
      period is much smaller in this case compared with intermediate
      $\lambda$. Accordingly, the phases behave sinusoidally for large
      $\lambda$, just as for the standard harmonic oscillator approached at
      small $\lambda$.}
    \label{Fig:Phases}
\end{figure}

As $\lambda \rightarrow \infty$, the stationary state passes through turning
points at a very high rate. The evolving state is obtained by concatenating
many branches of wave functions (\ref{psik1}) and (\ref{psik2}). The fact that
concatenations happen at different times for different energy eigenstates in a
superposition makes it hard to understand the long-term behavior of an
evolving state by analytic means. Nevertheless, numerical features, seen in
several model systems, show surprising simplifications, as demonstrated by the
example of a harmonic-oscillator system Hamiltonian in Figs.~\ref{Fig:psilam}
and \ref{Fig:PSLambda1}.  According to numerical results, each stationary
state as a function of $\tau$ is again governed by a sinusoidally varying
phase, except for small intervals around turning points. The main visible
difference with standard evolution is that the sinusoidal frequency is a
multiple of the expected frequency at $\lambda =0$ by a factor of
$\frac{1}{4}\pi$. An example is shown in Fig.~\ref{Fig:Phases}, where
movements of the clock through turning points (about two per system cycle) are
clearly visible for intermediate $\lambda$, while the many turning points the
clock goes through per system cycle for large $\lambda$ merely rescale the
system period but do not lead to noticeable deviations from sinusoidal
behavior.

\begin{figure}
    \centering
    \includegraphics[width=0.8\textwidth]{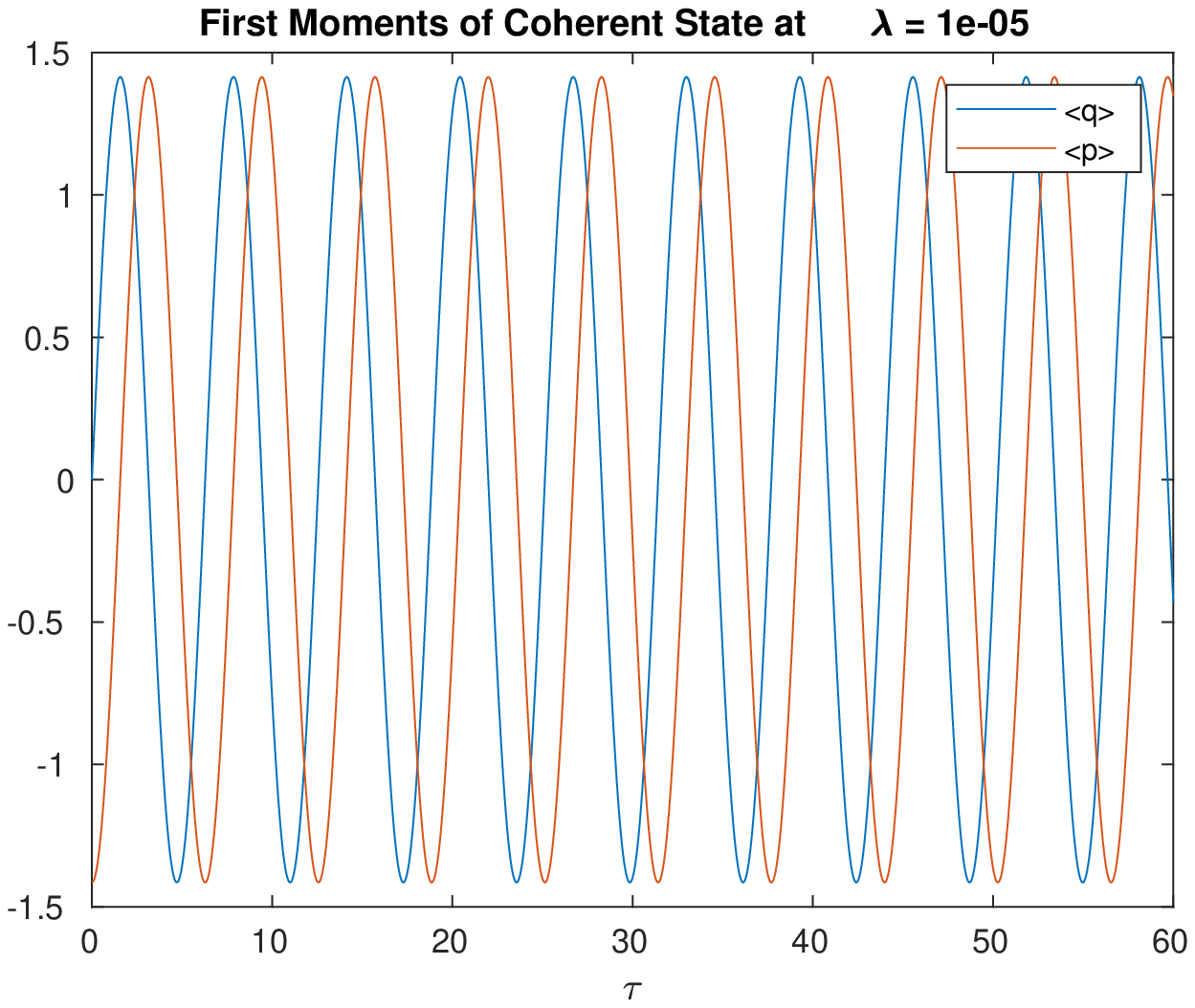}\\
\includegraphics[width=0.8\textwidth]{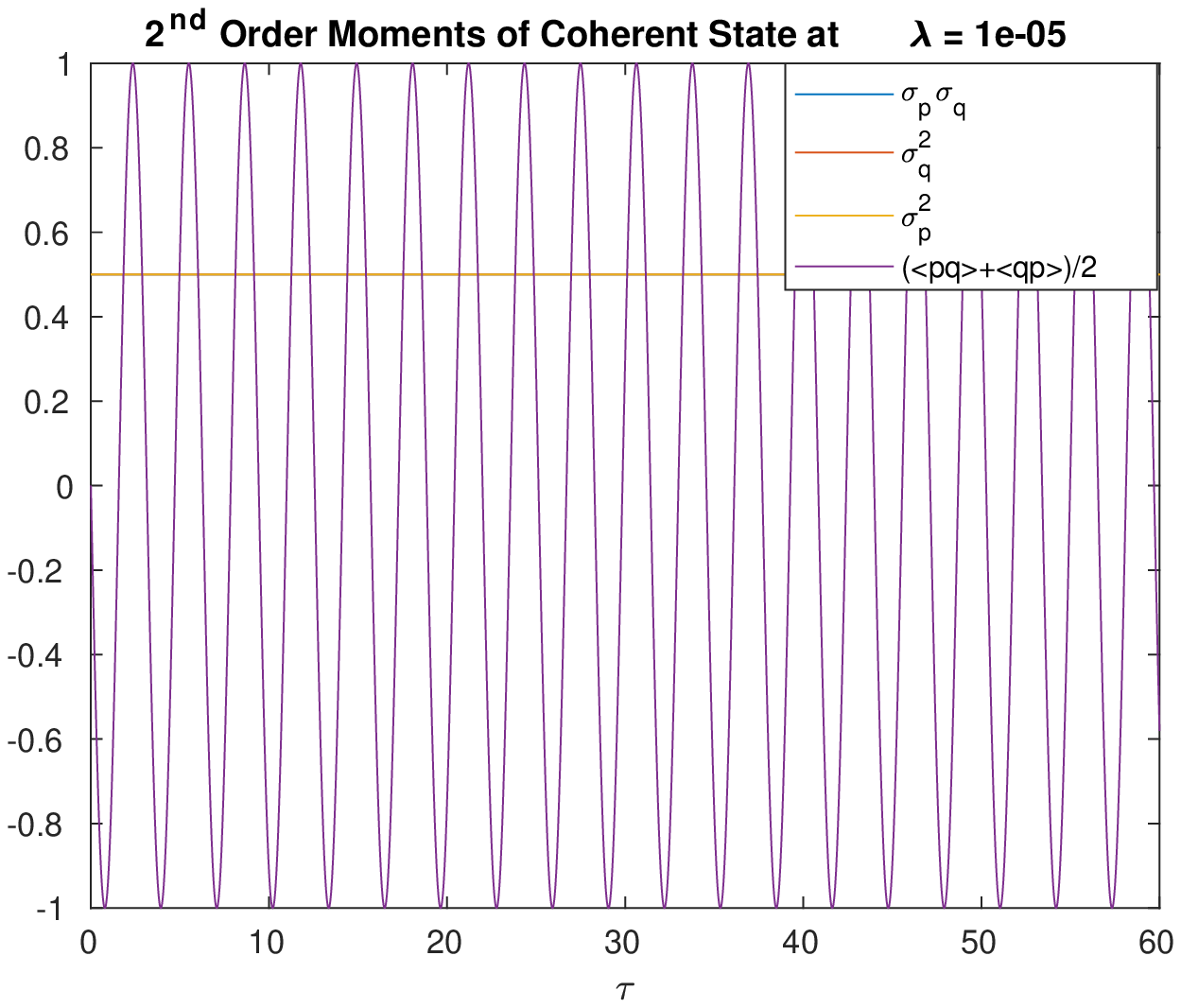}
\caption{Basic expectation values (top) and second-order moments (bottom) for
  small $\lambda$, using a harmonic system Hamiltonian. \label{Fig:SmallFirst}
}
\end{figure}

\begin{figure}
    \centering
    \includegraphics[width=0.8\textwidth]{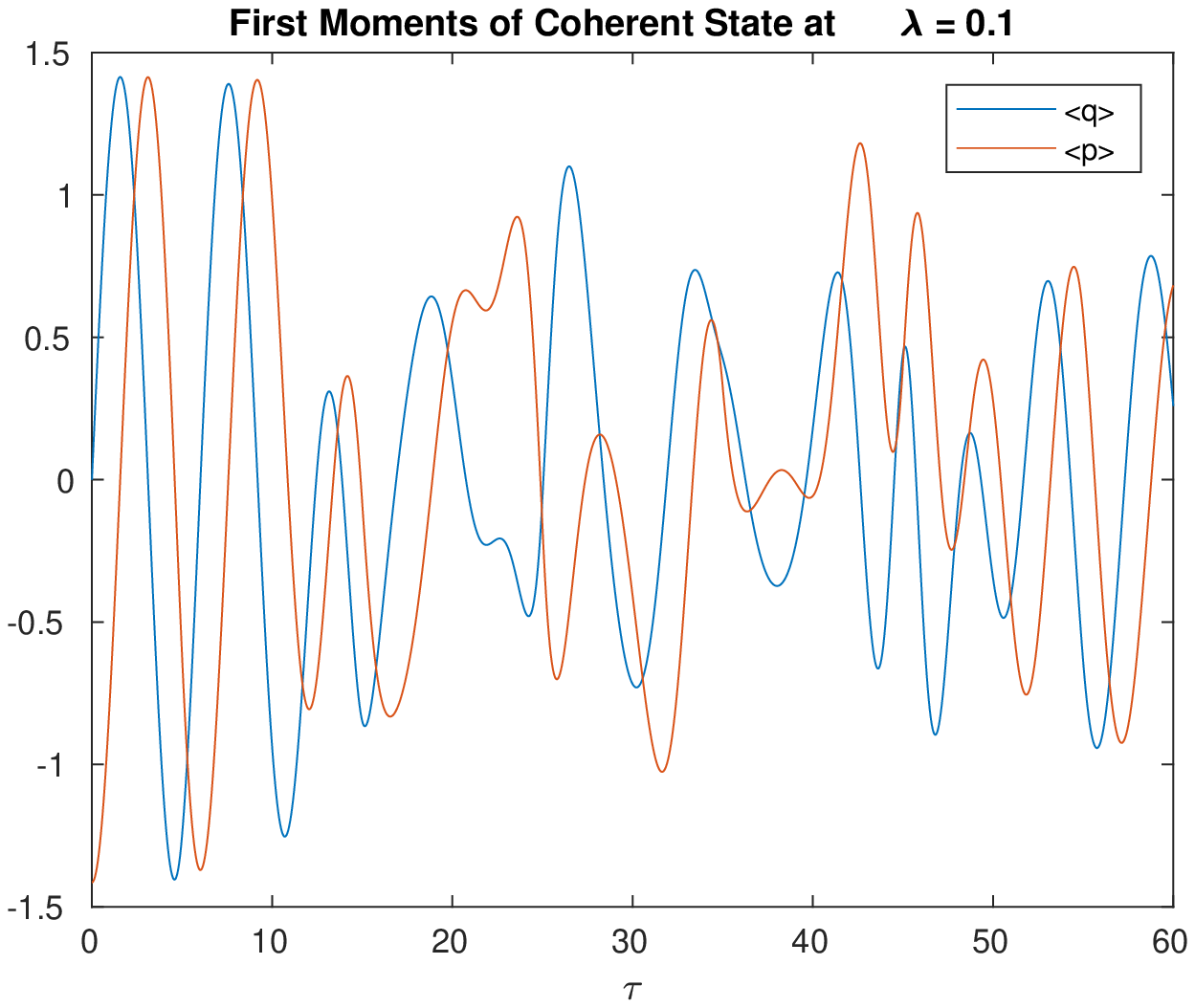}\\
    \includegraphics[width=0.8\textwidth]{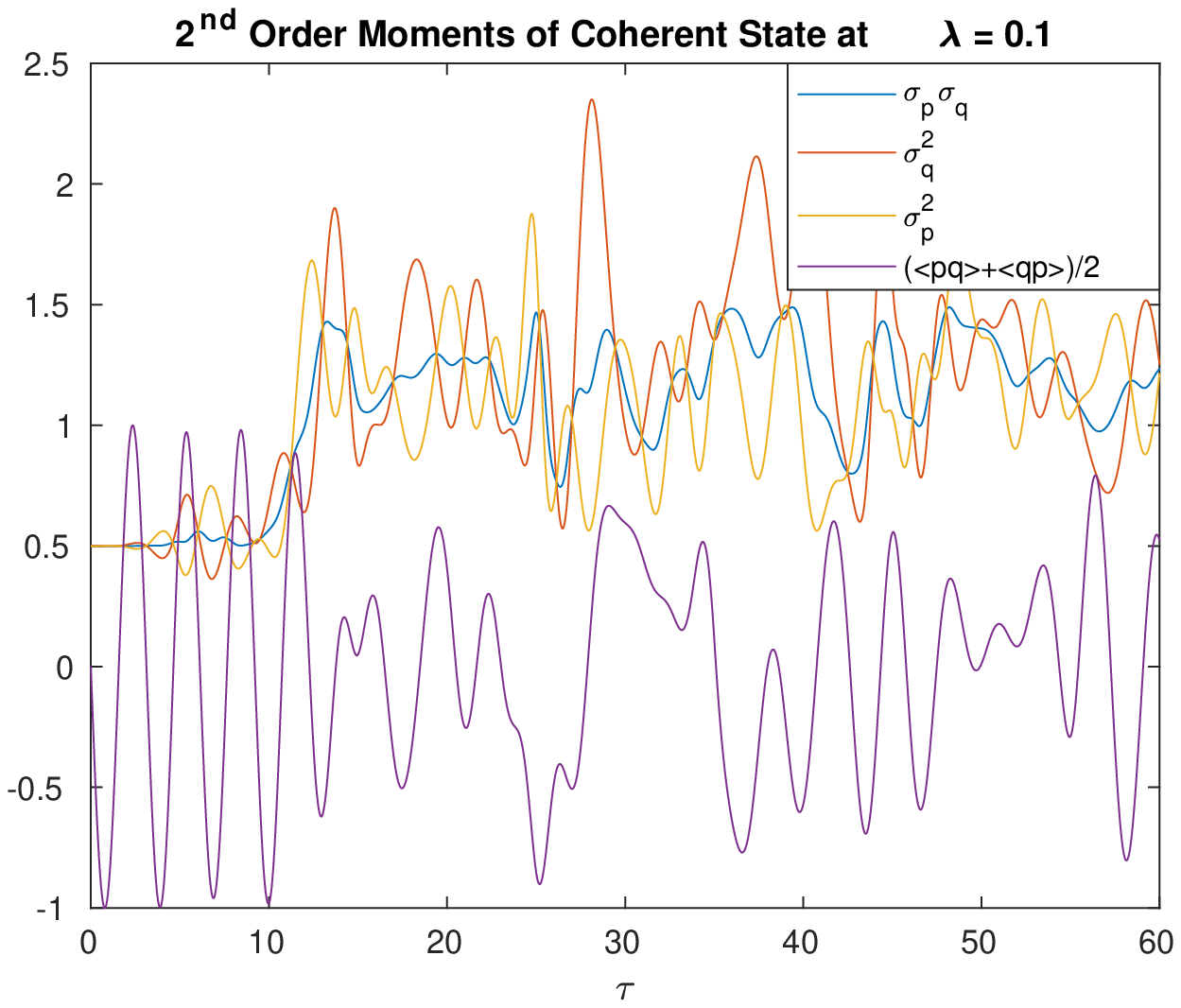}
    \caption{Basic expectation values (top) and second-order moments (bottom)
      for
  intermediate $\lambda$, using a harmonic system
  Hamiltonian. \label{Fig:MedFirst} }
\end{figure}

\begin{figure}
    \centering
    \includegraphics[width=0.8\textwidth]{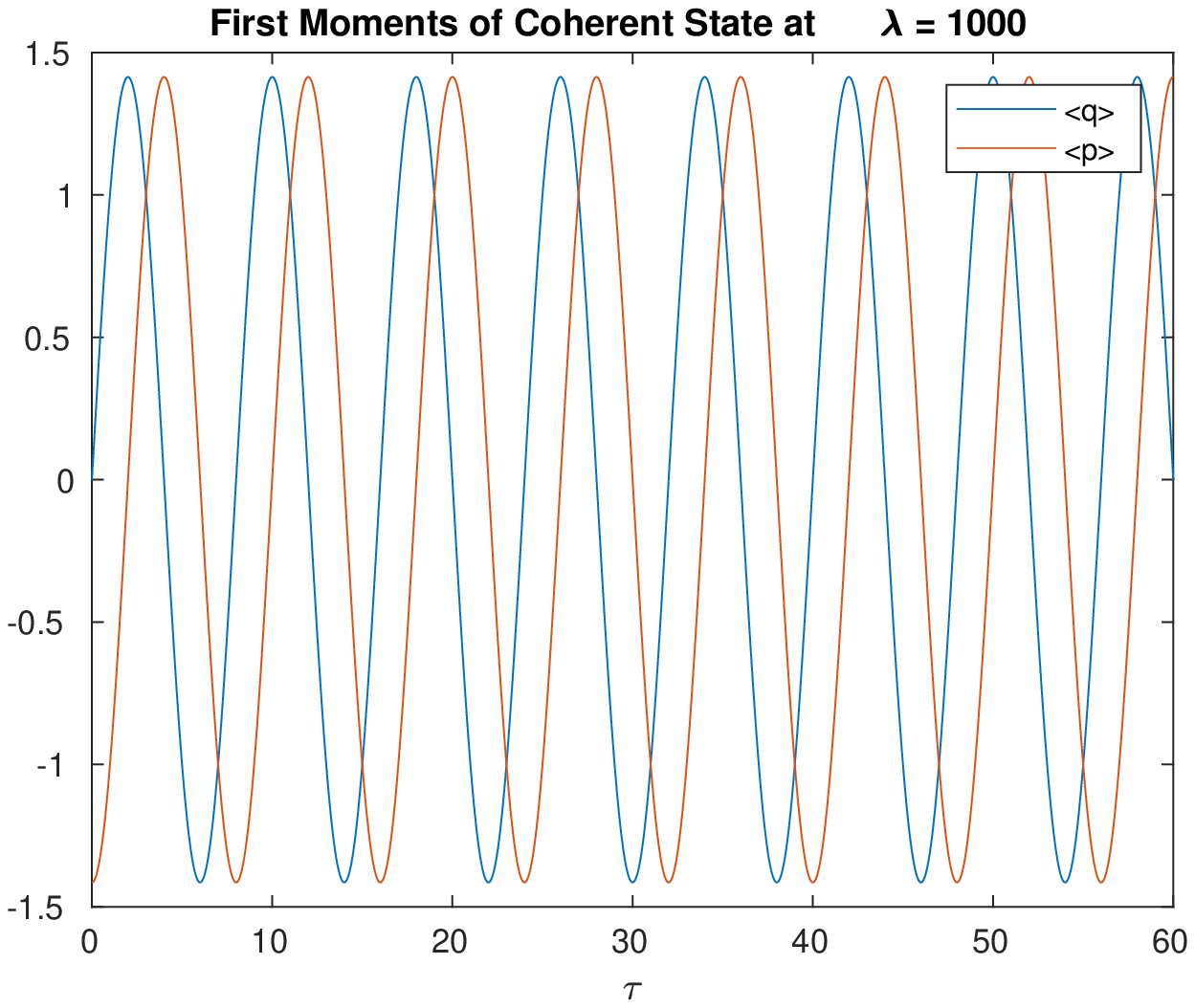}\\
    \includegraphics[width=0.8\textwidth]{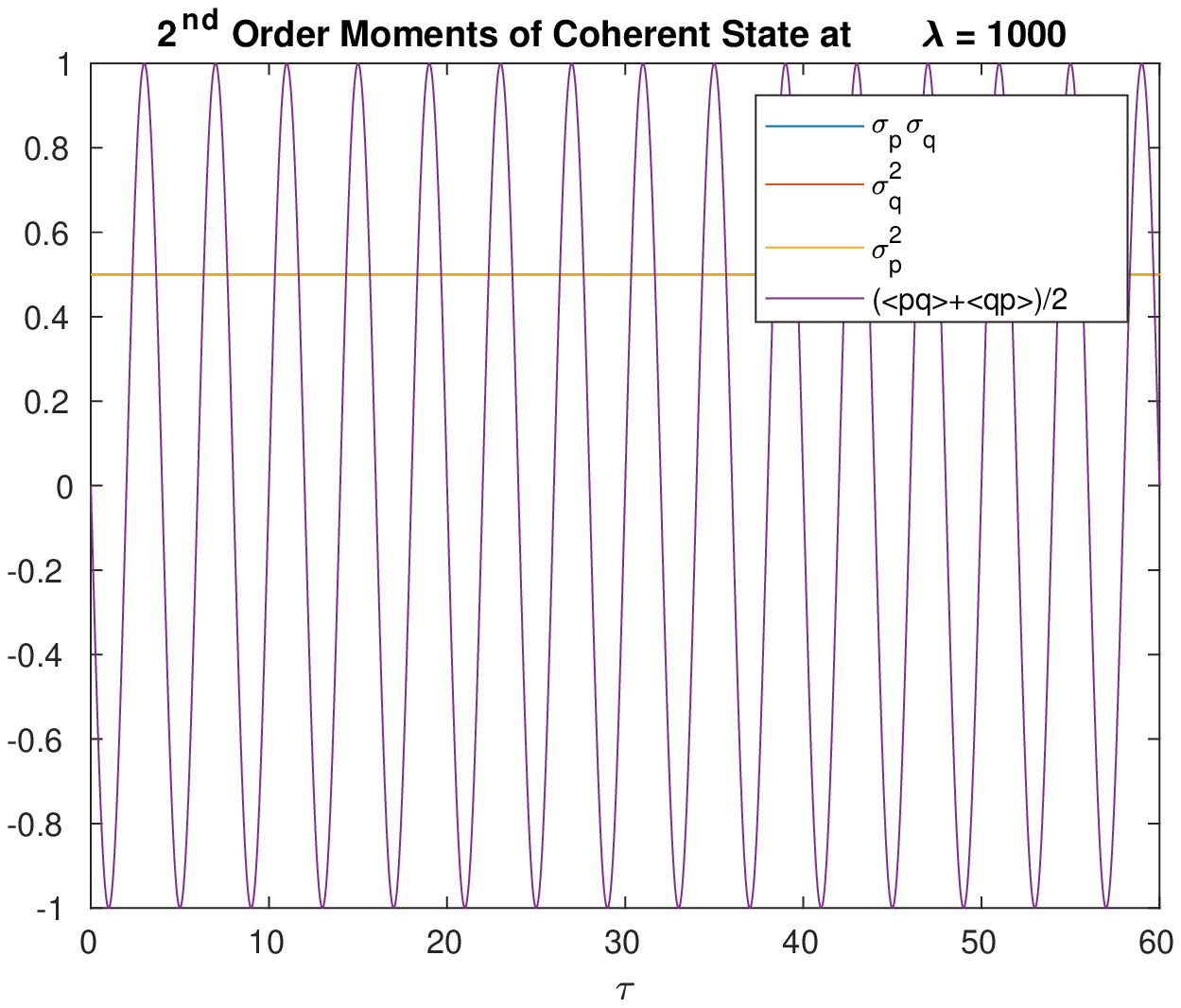}
    \caption{Basic expectation values  (top) and second-order moments (bottom)
      for large $\lambda$, using a harmonic system
  Hamiltonian. \label{Fig:LargeFirst} }
\end{figure}

The same features are shown by numerical evolution of expectation values of
$\hat{x}$ and $\hat{p}$ and their second-order moments (fluctuations and the
covariance), seen in Fig.~\ref{Fig:SmallFirst} for small $\lambda$ (close to
the standard harmonic oscillator), Fig.~\ref{Fig:MedFirst} for intermediate
$\lambda$, and Fig.~\ref{Fig:LargeFirst} for large $\lambda$. A single
stationary state does not show time-dependent expectation values and
moments. For this analysis, we have therefore chosen an initial state of
minimum uncertainty which would be a dynamical coherent state of the harmonic
oscillator ($\lambda=0$ or small).

For intermediate $\lambda$, the system rapidly loses coherence, as expected
because the $\phi$-Hamiltonian $\sqrt{H^2-\lambda^2\phi^2}$ is no longer
harmonic. As noted before, the phase of a wave function, obtained from
concatenated evolutions for each half-cycle of the clock, is not smooth. For
large $\lambda$, the high frequency at which the system crosses turning points
smoothes out the phase on scales larger than the clock period, which may have
been expected.  More surprisingly, the same non-harmonic Hamiltonian
$\sqrt{H^2-\lambda^2\phi^2}$ that implies rapid loss of coherence at
intermediate $\lambda$ leads to strongly coherent behavior at large
$\lambda$. This feature, as well as its explanation and applications below,
are the main results of \cite{LocalTime}, for which the present paper provides
a detailed discussion.

To explain this behavior, the large-$\lambda$ limit of Eqs.~(\ref{psik1}) and
(\ref{psik2}) can be simplified by ignoring the last exponential factors as
they approach zero in this limit: The first term in each of the last
exponentials is reduced to zero as $\lambda$ increases since the amplitude of
$\phi(\tau)$ approaches zero as $\lambda \rightarrow \infty$.  The second term
in each of the last exponentials is an arcsine function divided by $\lambda$,
which also approaches zero as $\lambda \rightarrow \infty$.  The reduced
equations therefore become
\begin{equation}\label{psik1large}
         \psi_k(\tau) \approx \psi_k(0)
        \exp \left( -\frac{i  (n+1/4)\pi E_k^2}{\lambda \hbar} \right)
\end{equation}
if $4n-1 \leq \tau/\phi_{\rm t} \leq 4n+1$, and
\begin{equation}\label{psik2large}
         \psi_k(\tau) \approx \psi_k(0)
        \exp \left( -\frac{i (n+3/4)\pi E_k^2}{\lambda \hbar} \right)  
\end{equation}
otherwise. Since $\phi_{\rm t}\to 0$ in this limit for fixed $E_k$, even small
changes in $\tau$ imply transitions between different clock cycles. Any
extended range of $\tau$ therefore leads to large numbers of clock cycles,
$n$, such that the remaining exponentials in (\ref{psik1large}) and
(\ref{psik2large}) are non-trivial even for large $\lambda$. However, there is
a negligible difference beteen $n+3/4$ and $n+1/4$ in the exponents, which
therefore are nearly identical in this limit.

As $\lambda \rightarrow \infty$ it is possible to approximate the floor
function in (\ref{n}), defining $n$, in a continuous form:
\begin{equation} \label{napprox}
        \frac{n}{\lambda} = \frac{\left \lfloor 1/4+\lambda \tau/(4E_k)
            \right \rfloor}{\lambda}  
         \approx \frac{\tau}{4E_k}\,.
\end{equation}
Using this result in Eq.~(\ref{psik1large}) yields
\begin{equation}
         \psi_k(\tau) \approx \psi_k(0)
        \exp \left(- \frac{i \pi  E_k}{4 \hbar} \tau \right) \,.
\end{equation}
Comparing this time-dependent phase with the usual solution
\begin{equation}
 \psi_k(t)= \psi_k(0) \exp(-iE_kt/\hbar)
\end{equation}
of the time-independent Schr\"odinger equation shows that the frequency of the
phase in the large-lambda limit is $\frac{1}{4}\pi$ times the frequency of the
system for $\lambda\to0$, in agreement with our numerical plots for the
harmonic oscillator with Hamiltonian
$\hat{H}=\frac{1}{2}(\hat{p}^2+\hat{x}^2)$: When $\lambda = 10^{-5} $
(approximating the limit $\lambda \rightarrow 0$), a sinusoidal function with
a period of $2\pi$ is returned, as expected for the ground state of the
harmonic oscillator of frequency parameter $\omega=1$.  For large values of
lambda, $\lambda = 10^3$ (approximating $\lambda \rightarrow \infty $), the
period of the oscillations is multiplied by a factor of $4/\pi$ resulting in a
value of $8$. With intermediate values, turning points are spaced at easily
visible intervals, for instance located at $\tau = 5 + 10j$ with integer $j$
in the case of $\lambda = 10^{-1}$ as shown in Fig.~\ref{Fig:Phases}.

\subsection{Non-harmonic examples}
\label{s:NonHarm}

Our specific equations for the $\tau$-dependent phase can be used for any
system Hamiltonian. The harmonic example enjoys the most coherent dynamics and
therefore highlights any loss of coherence implied by an oscillating
fundamental clock. Non-harmonic systems cannot regain strong coherence for
large $\lambda$ simply because their dynamics is not coherent, but it is
nevertheless possible to see an approach to standard quantum mechanics for a
periodic clock with large $\lambda$.

\begin{figure}
    \centering
    \includegraphics[width=0.8\textwidth]{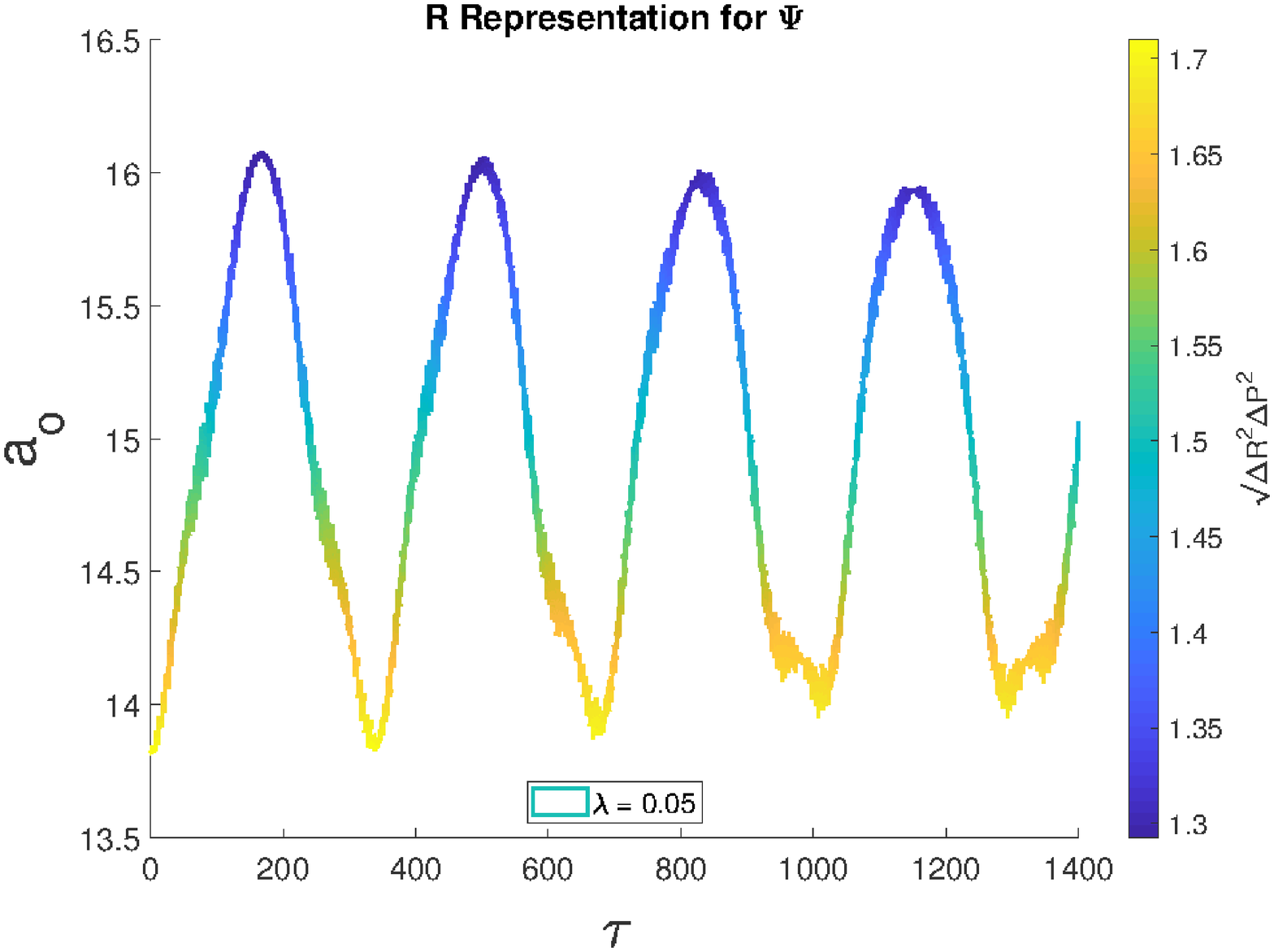}\\
    \includegraphics[width=0.8\textwidth]{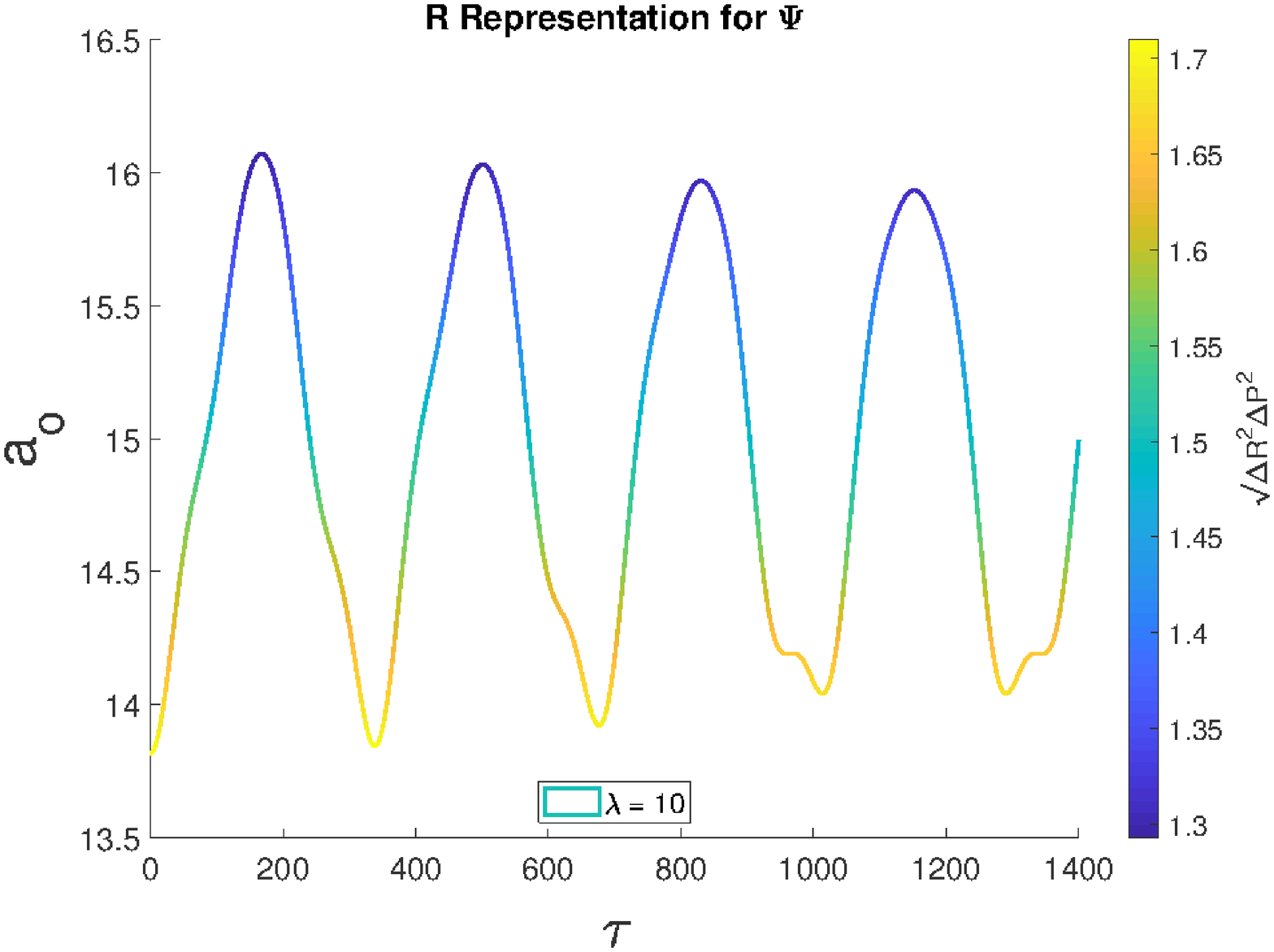}
    \caption{Expectation value and quantum uncertainty of the orbital radius
      with a standard hydrogen Hamiltonian. The quantum uncertainty is similar
      for all $\lambda$, while the time-dependent expectation value has an
      additional high-frequency contribution for intermediate $\lambda$ (top)
      compared with large $\lambda$ (bottom) or standard quantum
      mechanics. \label{Fig:Hydro} }
\end{figure}

Figure~\ref{Fig:Hydro} shows the example of a basic hydrogen Hamiltonian,
based on the Coulomb potential, with a non-coherent initial state, chosen as a
certain superposition of finitely many energy eigenstates. Quantum
fluctuations vary rather strongly for any $\lambda$, as they would also do in
standard quantum mechanics in this case. There is a notable difference in the
expectation values for different $\lambda$, shown by a high-frequency signal
visibly superimposed for intermediate $\lambda$ that disappears for large
$\lambda$ where the expectation values approach the standard behavior.

Similar features are obtained for a non-harmonic clock Hamiltonian. The main
difference of a non-harmonic system compared with a harmonic clock is that its
period depends on the energy. However, in our coupled system the clock period
$T_{\rm C}=4\phi_{\rm t}=4 E_k/\lambda$ already depends on the system energy,
which equals the clock energy by the energy-balance constraint. Non-harmonic
behavior of the clock therefore does not introduce crucial new features,
although it makes explicit calculations more complicated.

\section{Observational bounds}

So far, we have found one quantitative difference between the evolution of a
harmonic system with respect to an oscillating clock one one hand, and with
respect to an absolute time on the other: A rescaling of the system period by
a factor of $4/\pi$ for the clock Hamiltonian used here, as described in
Section~\ref{s:Largel}. However, this difference is not observable because it
depends only on the clock Hamiltonian and therefore rescales all system
frequencies or their characteristic time scales in the same way. The rescaling
factor can therefore be removed by absorbing it in a bare frequency
$\omega'= 4 \omega/\pi$ that appears in the mathematical expression of the
system Hamiltonian, and therefore gives rise to the observed frequency
$\omega$ after our rescaling.  The scaling factor depends neither on the
energy or initial state for a given system, nor on the system itself. It
depends only on the clock dynamics, which is the same for all systems if the
clock is fundamental.

The second implication of an oscillating clock is its effect on the
decoherence time, depending on the clock parameter $\lambda$. Since a coherent
state of the harmonic oscillator has an infinite decoherence time in standard
quantum mechanics, a finite decoherence time as shown by our numerical results
cannot be eliminated by a simple rescaling.  Moreover, the time of how long a
system can maintain coherence is under good observational control. For
instance, the current relative precision of $10^{-19}$ of atomic clocks
\cite{LatticeClock} could not be obtained if nature had provided us with an
intermediate $\lambda$ in our fundamental clock that would destroy coherence
or a stable system period over many system cycles.

\begin{figure}
    \centering
    \includegraphics[width=0.8\textwidth]{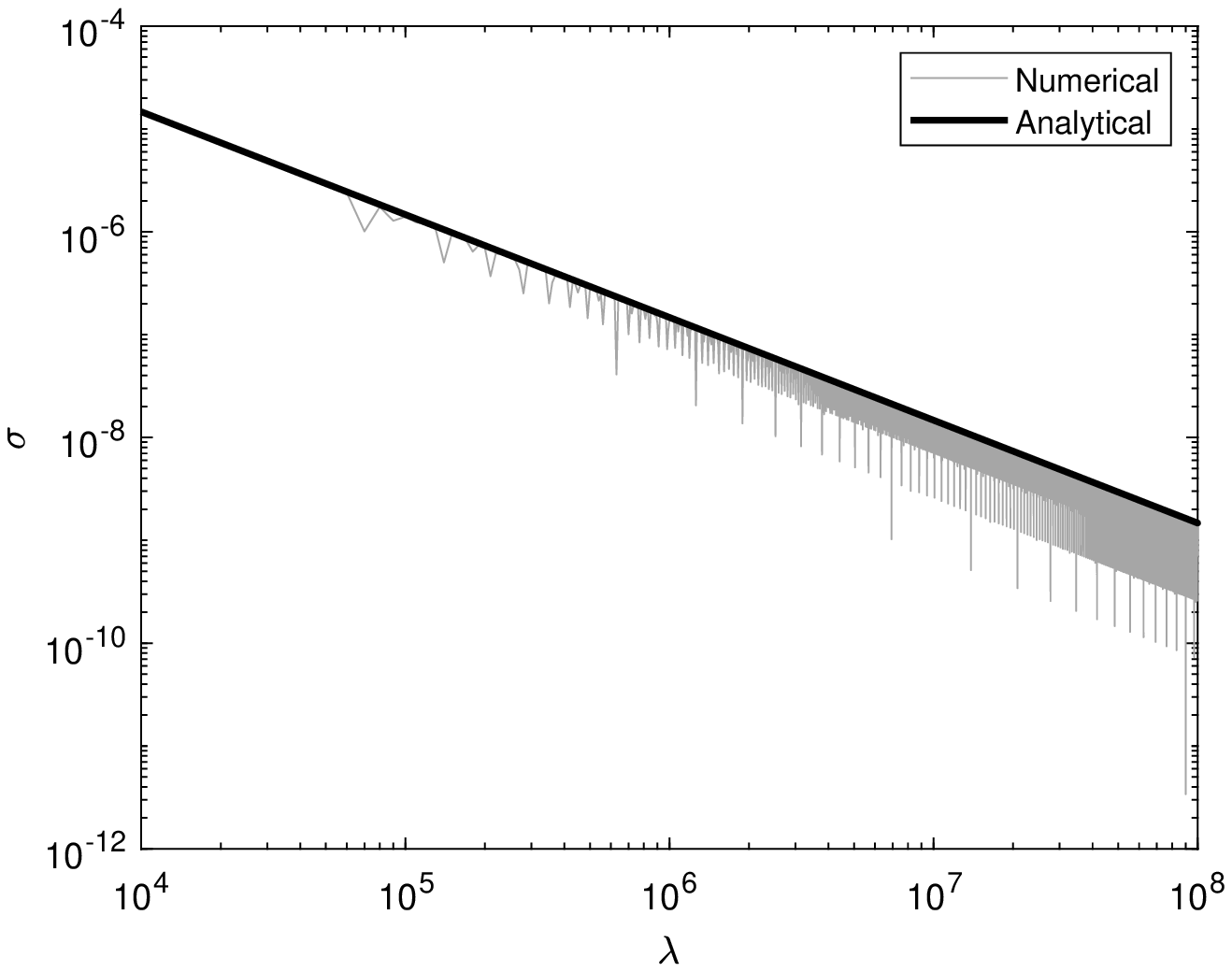}
    \caption{Relative standard deviation of the system period over many system
      cycles as a function of $\lambda$. The analytical approximation
      (\ref{sigma}) agrees well with a numerical computation of many system
      periods, both confirming a $1/\lambda$-behavior. The analytical result
      is rescaled by a factor of $2/\pi$ to take into account the average of a
      sine fuction over a quarter-cycle.  \label{Fig:AnalyticalvsNumerical}}
\end{figure}

Taking (\ref{napprox}) to the next order in $1/\lambda$ indicates that
relative deviations from strict sinusoidal behavior should be of this
order. More precisely, for given finite $\lambda$, we can compute the variance
of the phase $\Theta_k(\phi(\tau))$, given in (\ref{Theta}) around
\begin{equation}
 \Theta_k^{\infty}(\tau):=\lim_{\lambda\to\infty}\Theta_k(\phi(\tau))=
 \frac{\pi}{4}E_k\tau
\end{equation}
averaged over a quarter-cycle of $\phi$:
\begin{eqnarray}
 \sigma^2 &=& \frac{1}{\phi_{\rm t}} \int_0^{\phi_{\rm t}}
   \left(\Theta_k(\phi(\tau))- \Theta_k^{\infty}(\tau)\right)^2{\rm
     d}\tau\nonumber\\
&=& \frac{E_k^4 (21\pi^2-1024/5)}{24^2\lambda^2\hbar^2} \,. \label{sigma}
\end{eqnarray}
Introducing the clock period $T_{\rm C}=4\phi_{\rm t}=4E_k/\lambda$ and the system
period $T_{\rm S}=2\pi\hbar/E_k$ allows us to eliminate the parameters
$\lambda$ and $E_k$ in favor of more general clock
characteristics. Equation~(\ref{sigma}) then takes the form
\begin{equation} \label{TT}
 T_{\rm C}= \frac{48 \sigma T_{\rm S}}{\pi\sqrt{21\pi^2-1024/5}}\approx 9.7
 \sigma  T_{\rm S}\,.
\end{equation}
The dependence on $\sigma$ implies a strong magnification factor that can make
precision measurements sensitive to a small period of the fundamental clock
even if they operate on a larger system period. Indirect bounds on the clock
period therefore come within reach.

Even though (\ref{TT}) no longer refers directly to an energy eigenvalue, its
derivation remains strictly valid only for a system period of  an eigenstate. It
is more difficult to obtain analytical control over the variance of the system
period of a superposition of several energy eigenstates, such as a coherent
state. A natural expectation is that $E_k$ in (\ref{sigma}) should then be
replaced with a certain expectation value determined by the Hamilton operator
$\hat{H}$, for which there are different options. As shown by
Fig.~\ref{Fig:AnalyticalvsNumerical}, replacing $E_k^2$ in (\ref{sigma}) with
the expectation value $\langle\hat{H}^2\rangle$ is in good agreement with
numerical computations of the relative standard deviation, $\sigma$, of the
system period over many system cycles, shown as a function of $\lambda$ in the
figure. While the close agreement is encouraging, it is also surprising and
remains incompletely understood: The quarter-cycle calculation in
(\ref{sigma}) does not directly refer to turning points of the clock and is
therefore insensitive to flipping the signs of the phase according to
(\ref{Theta1}) and (\ref{Theta2}), while the latter is important for global
evolution over many cycles during which we notice the restoration of
coherence. The quarter-cycle calculation does, however, depend on the
non-linearity of the phase implied by an oscillating clock, which is most
prominent in the approach to a turning point.

For very large $\lambda$, one should evolve through many system cycles in
order to determine the standard deviation accurately, considering the
collection of cycles as a statistical ensemble for the observable period. Such
long-term evolution was beyond the numerical capacity available to
us. However, since the upper bound on the standard deviation shown in the
figure follows a simple $1/\lambda$-behavior, we are justified in
extrapolating it to values of $\lambda$ even larger than those shown in the
plot.  According to this extrapolation to large $\lambda$, a relative accuracy
of $10^{-19}$, as reported for recent atomic clocks in \cite{LatticeClock},
requires ten orders of magnitude less than the smallest relative standard
deviations shown in Fig.~\ref{Fig:AnalyticalvsNumerical}. Since one order of
magnitude in $\lambda$ corresponds to about one order of magnitude in
$\sigma$, taking into account the numerical factor of about ten in (\ref{TT}),
we need $\lambda$ at least as large as $10^{18}$ times the system frequency,
or a fundamental clock period of at most $10^{-18}$ times the system period.

The system period (or atomic clock period) used in \cite{LatticeClock} is
based on the transition from the $^3P_0$ state to the $^1S_0$ ground state of
Strontium, with a wave length of $698\,{\rm nm}$, amounting to a system period
of about $2\,{\rm fs}$. The upper bound on the fundamental clock frequency is
therefore
\begin{equation}\label{Upper}
 T_{\rm C}<10^{-18}\cdot 2\,{\rm fs}= 2\times 10^{-33}\, {\rm s}\,.
\end{equation}
Although this upper bound is still several orders of magnitude larger than the
Planck time $t_{\rm P}= 5 \times 10^{-44}\,{\rm s}$, which is often suggested
as a fundamental period of time based on dimensional arguments, it is much
smaller than any value that could at present be obtained from a direct time
measurement. For instance, the current value of the shortest time interval
measured directly, given by the photon travel time $247\cdot 10^{-21}{\rm s}$
across a hydrogen molecule \cite{Zepto}, is more than ten orders of magnitude
larger than our upper bound.

Our new upper bound is also stronger than previous indirect measurements of
short time scales.  In particular, the shortest length measurements currently
possible are of the order $10^{-19}\,{\rm m}$, achieved at high-energy
particle accelerators. Using the speed of light, this value translates into an
upper bound of $10^{-19}\,{\rm m}/c\approx 3\cdot 10^{-28}\,{\rm
  s}$ on the time scale, about five orders of magnitude above our new upper
bound. This value, like ours, is based on an indirect measurement because it
translates a direct measurement of a scattering cross section into a length,
and then into a time parameter. By exploiting the dephasing time, our indirect
measurement is much more sensitive even than indirect measurements at high
energy.

\section{Conclusions}

We have analyzed a combination of an oscillating clock variable $\phi$ and an
evolving system degree of freedom $x$, coupled minimally through an
energy-balance constraint (\ref{C4}). Expressed as relational evolution of $x$
with respect to $\phi$, the dynamics is governed by a standard Schr\"odinger
equation (\ref{SchroedingerE}) with time-dependent Hamiltonian
$\sqrt{\hat{H}^2 -\lambda^2\phi^2}$ if $\hat{H}$ is the Hamiltonian that
determines the energy of the system. Therefore, an oscillating clock implies
that the usual equality between the energy operator
$i\hbar\partial/\partial t$ and the system Hamiltonian no longer holds. The
operator $i\hbar\partial/\partial \phi$ instead determines the momentum of the
clock variable $\phi$ and therefore its kinetic energy, but not the full clock
energy because an oscillating clock also has potential energy. The
energy-balance constraint makes sure that the combined energy of the system
and the clock is conserved. But the clock momentum, measured in quantum
mechanics by $i\hbar\partial/\partial \phi$, no longer equals the system
Hamiltonian.

Experience with standard quantum mechanics would suggest that evolution with a
Hamiltonian $\sqrt{\hat{H}^2 -\lambda^2\phi^2}$ is rather complicated for any
$\lambda\not=0$, even if $\hat{H}$ belongs to the harmonic oscillator as in
our main example. (Fractional powers of Hamiltonians can imply additional
subtleties; see for instance \cite{FractionalHarmonic}.) This expectation is
confirmed by Fig.~\ref{Fig:psilam} for small and intermediate values of
$\lambda$, defined such that $\lambda T_{\rm S}/\langle\hat{H}\rangle$ is not
very large, where $T_{\rm S}$ is the system period and the expectation value
$\langle\hat{H}\rangle$ of the system energy is taken in an initial system
state. For this range of $\lambda$, the coherence of an initial standard
coherent state is quickly lost as soon as the term $\lambda^2\phi^2$ becomes
relevant in the action of $\sqrt{\hat{H}^2 -\lambda^2\phi^2}$ on an evolving
state. For large $\lambda$, one would then expect that coherence is lost even
faster, well before the system can complete a single period or just move in a
noticeable way. Surprisingly, however, coherence is restored for very large
$\lambda$, as also shown in Fig.~\ref{Fig:psilam}.


The unexpected restoration of coherence for small periods of the fundamental
clock demonstrates that an oscillating fundamental clock is consistent not
only conceptually, as already shown in \cite{Gribov}, but also physically:
Even though the potential required for a periodic clock affects the coupling
between system and clock through the energy-balance constraint, leading to a
non-harmonic system Hamiltonian of the form
$\sqrt{\hat{H}^2-\lambda^2\phi^2}$, coherence can be maintained for
surprisingly long times for large $\lambda$. Here, it is important to note
that the dynamics is governed not only by the linear Schr\"odinger equation
(\ref{SchroedingerE}), but also by a discrete process given by flipping the
sign of the phase according to (\ref{Schroedingertau}). The precise
mathematical origin of restored coherence remains to be understood. This
restoration of coherence is not simply a perturbation of the standard harmonic
oscillator because it occurs for large $\lambda$, where perturbation theory
cannot be used in $\sqrt{\hat{H}^2-\lambda^2\phi^2}$. While intermediate
$\lambda$ are ruled out by observations of long coherence in isolated quantum
systems, large $\lambda$ and therefore sufficiently small fundamental periods
are consistent with current observations. A slight dephasing persists,
however, even at large $\lambda$, giving rise to our upper bound
(\ref{Upper}).

The origin of the coherence effect lies in quantum mechanics with an
oscillating fundamental clock. It does not have a complete classical analog. A
fundamental period of time would imply that a system period that is not an
integer multiple of the fundamental period cannot be sampled
precisely. Successive system periods may therefore appear slightly longer or
shorter depending on which system cycle an incomplete clock period is
attributed to in a measurement. This classical model would also lead to a
certain variance in system periods, but any such effect would quickly average
out over a few system cycles. Moreover, two systems starting at the same time
would be affected by the over/undercounting of complete clock cycles in the
same way. If they are synchronized, like an atom and a photon of the right
energy to generate a transition of energy levels or like two atomic clocks,
they would therefore not get out of tune by classical variations of the system
period. Observable implications of such a classical model would be
insignificant.

Our new quantum effect, which also implies variations of the system frequency,
is of a different nature. It acts on subtle coherence properties in a
superposition of energy eigenstates of the system. If, again, we have two
synchronized systems starting at the same time, they do not have identical
quantum states, and therefore are affected in different ways by the new
coherence and dephasing effects. Their system periods still vary, and
subsequent cycles of one system present a statistical ensemble independent of
the cycles of the other system because the initial states are largely
independent except in certain macroscopic properties that have been arranged
to agree in the synchronization procedure. Therefore, detuning can in
principle be observed by comparing the two systems. Similarly, if the two
systems are an atom and a photon which are ``synchronized'' in the sense that
the energy of the photon matches a transition energy of the atom, the atom and
photon states are necessarily different and therefore react differently to the
new coherence effect.

Unfortunately, it seems difficult to derive the new coherence effect at large
$\lambda$ in a controlled analytical approximation, even though it is clearly
presented by numerical simulations. We have been able to compute the variance
of the system period (\ref{sigma}) in good agreement with a numerical analysis
of the statistical ensemble given by the periods of a simulated wave function,
as demonstrated by Fig.~\ref{Fig:AnalyticalvsNumerical}. However, the
agreement remains somewhat mysterious because the calculation in (\ref{sigma})
is based on a quarter cycle of the fundamental clock, which is a much shorter
time than spanned by the large number of system periods that are used in the
numerical analysis. The agreement is encouraging and supports the relevance of
the new effect, but for a detailed analysis and further predictions, as well
as stricter upper bounds, it would be desirable to have an analytical
approximation that could accurately describe the statistical and coherence
properties of the system over many periods. Developing such an approximation
is challenging because the dynamics over many clock cycles is governed not
only by the linear partial differential equation (\ref{SchroedingerE}), but
also by the phase ``reflections'' in Eqs.~(\ref{Theta1}) and (\ref{Theta2}).

Our results have several conceptual implications for the quantum nature of
clocks and time. The fact that evolution for large $\lambda$ agrees with
$\lambda=0$ to a good degree demonstrates that deparameterization, a procedure
going back to Dirac and now widely used to evade the problem of time in
quantum gravity and quantum cosmology, can be considered a controlled
approximation of quantum dynamics at least as long as system periods are not
Planckian. The deparameterization procedure may therefore be applied at low curvature, but it
remains questionable at Planckian curvature where the system period (or any
rate of change if the system is not periodic, like the expanding universe)
itself is Planckian and compariable with the period of a fundamental
clock. New, unexpected effects may therefore be implied by a fundamental clock
at the big bang, which remain to be evaluated.

Finally, our analysis has shown that a periodic fundamental clock is, in
general, in a superposition of different clock cycles. We reach this
conclusion because the turning points of the clock variable, minimally coupled
to the system through energy balance, depend on the system energy as shown by
(\ref{phit}). Since a system is generically in a superposition of different
energy eigenstates, a fundamental period will quickly evolve into a
superposition of different clock cycles even if it is assumed to start in a
state that is sharply peaked around a given clock value. In particular, an
application of equation (\ref{phitau}) shows that the requirement of having a
unique value of global time $\tau$ for all energy eigenstates in superposition
implies different clock values $\phi_k(\tau)$ for different energy eigenvalues
$E_k$. The more time $\tau$ progresses, the more the various $\phi_k$ for a
given system state differ, potentially stretching over many clock cycles. Our
procedure of concatenating half-cycle evolutions for each eigenstate and then
bringing them back into superposition allows us to avoid dealing directly
with a complicated clock state, but such a state is indirectly realized.
This effect does not appear in non-periodic clocks such as those used in
deparameterization. To the best of our knowledge, our model is the first in
which the clock is truly quantum, understood in the sense that it is by
necessity in a superposition of different cycles.

\section*{Acknowledgements}

We thank Ding Ding, Bianca Dittrich, Kurt Gibble and Philipp Hoehn for
discussions.  This work was supported in part by NSF grant PHY-1912168. LM was
supported by a Gates Scholarship.

\begin{appendix}

\section{Ingredients of the code}

We present crucial parts of the MATLAB program used for our results as
pseudocode:
\begin{itemize}
\item Calculating $\phi(\tau)$:  
  \begin{verbatim}
if ((tau/Phi_t) <= 4*n+1)
  Phi_tau = t-4*n*Phi_t;
elseif ((tau/Phi_t) > (4*n+1))
  Phi_tau = (4.*n+2)*Phi_t-t;
  \end{verbatim}
\item Calculating $n$:
  \begin{verbatim}
n = floor((1+(tau/Phi_t))/4);
  \end{verbatim}
\item Calculating $\Theta(\phi(\tau))$:
  \begin{verbatim}
if ((tau/Phi_t)>(4*n+1))
  Theta = exp(pi*i*(n+1/2)*((k+1/2)^2)/lambda/hbar)*
    exp(-i/2/lambda/hbar*(lambda*Phi_tau*((k+1/2)^2-lambda^2*Phi_tau^2)^.5))*
    exp(-i/2/lambda/hbar*(k+1/2)^2*asin(lambda*Phi_tau/(k+1/2)));
elseif ((tau/Phi_t(k,lambda))<=(4*n(k,lambda,tau)+1))
  Theta = exp(pi*1i*n*(k+1/2)^2/lambda/hbar)*
    exp(i/2/lambda/hbar*(lambda*Phi_tau*((k+1/2)^2-lambda^2*Phi_tau^2)^.5))*
    exp(i/2/lambda/hbar*(k+1/2)^2*asin(lambda*Phi_tau/(k+1/2)));
end
\end{verbatim}
\item Calculating zero crossing statistics:
  \begin{verbatim}
for (a given range of lambda)
  Qbar(t)=Integral(|Psi(q,t)|^2*q)
  for (a given number of q intervals)
    distance_from_exact= (lambda is infty zero) - findzero(Qbar,q interval)
    std_dev(distance_from_exact)
    average(distance_from_exact)
  \end{verbatim}
\end{itemize}
    
\end{appendix}


\end{document}